\documentclass[12pt]{article}
\usepackage[utf8]{inputenc}

\usepackage{graphics}
\usepackage{epstopdf}
\usepackage{csquotes}
\usepackage{psfrag}
\usepackage{amsmath}
\usepackage{amsfonts}
\usepackage{float}
\usepackage{lineno}
\usepackage{rotating}
\usepackage{url}
\usepackage{xcolor}
\usepackage[section]{placeins}
\usepackage[a4paper, total={6.5in, 10in}]{geometry}

\usepackage[english]{babel}

\newcommand{\benum}{\begin{enumerate}}
\newcommand{\eenum}{\end{enumerate}}
\newcommand{\bize}{\begin{itemize}}
\newcommand{\eize}{\end{itemize}}
\newcommand{\eeq}{\end{equation}}
\newcommand{\beqa}{\begin{eqnarray}}
\newcommand{\eeqa}{\end{eqnarray}}
\newcommand{\beqsub}{\begin{subequations}}
\newcommand{\eeqsub}{\end{subequations}}

\definecolor{red}{rgb}{1,0,0} 
\definecolor{blue}{rgb}{0,0,0.8} 
\definecolor{green}{rgb}{0,0.5,0}

\title{
Cancer model with moving extinction threshold reproduces real cancer data
}

\author{
Frank Bastian\footnote{University College Cork, School of Mathematical Sciences, Western Road, Cork, T12 XF62, Ireland}, Hassan Alkhayuon$^*$, Kieran Mulchrone$^*$, \\Micheal O'Riordain\footnote{Department of Surgery, Mercy University Hospital, Cork, T12 WE28, Ireland}, and Sebastian Wieczorek$^*$}

\date{\today}

\begin{document}

\maketitle

\begin{abstract}

We propose a simple dynamic model of cancer development that captures carcinogenesis and subsequent cancer progression. A central idea of the model is to include the immune system as an extinction threshold, similar to the strong Allee effect in population biology. 
We first identify the limitations of commonly used Allee effect models 
in reproducing typical cancer progression. We then address these limitations by deriving a new model that incorporates:
(i) random mutations of stem cells at a rate that increases with age and 
(ii) immune response whose strength may also 
vary over time.

Our model accurately reproduces a wide range of real-world cancer data: 
the typical age-specific cumulative risk of most human cancers, the progression of breast cancer in mice, and the unusual age-specific cumulative risk of breast cancer in women. In the last case, we use a moving extinction threshold to reflect the different immune response at different phases of the menstrual cycle and menopausal treatment. This provides new insights into the effects of hormone replacement therapy and menstrual cycle length. 
This moving threshold approach can be applied to a variety of other cancer scenarios where the immune response or other important factors may vary over time.

\end{abstract}

\section{Introduction}

Cancer is one of the leading causes of death world wide, accounting for 10 million fatalities a year~\cite{cost_cancer}.
The global lifetime risk of developing any type of cancer is estimated to be 
approximately $25\%$, however, this varies significantly with socioeconomic factors encapsulated in the human development index~\cite{risk_cancer_HDI}.
Additionally, cancer cases are forecast to rise by 77\% between 2022 and 2050~\cite{WHO_cancer_grow} due to several factors including changing demographic profiles in developed countries and improving living standards in underdeveloped countries.  The spectre of an increasing cancer burden is a cause for concern in both economic and social terms~\cite{cost_cancer,social_impact_cancer}.
There is an urgent need to  better understand the critical points of cancer development and to improve cancer treatment~\cite{challenges_cancer_treatment}.  This is evidenced by several initiatives such as the war on cancer in the USA~\cite{war_on_cancer}, or Europe's beating cancer plan~\cite{EU_plan}.

Among the different  types of cancer, breast cancer accounts for $12.5\%$ of all new annual cancer cases worldwide, making it  one of the most common cancers in the world~\cite{breast_org}. 
 In the USA, for example, $13.1\%$ of women and less than $0.14\%$ of men will develop breast cancer in their lifetime~\cite{seer_female_breast_fact_US,acs_male_breast_fact_US}.
%
Moreover, we observe that for breast cancer in women, the age-specific cumulative risk 
appears to increase polynomially up
to the age of menopause and linearly thereafter;\footnote{Linear increase does not accelerate, polynomial accelerates and exponential accelerates faster than polynomial.} see also~\cite{epidemiology_breast_cancer,Clemmesen_hook}.
This is quite unusual for cancer and raises important research questions about the dynamic mechanisms underlying the development of different cancers.

Over the last century several effective treatment therapies have been developed e.g. surgery, chemotherapy and radiation~\cite{chemo_radio}.
These treatments are often applied aggressively, i.e. administering a {\em maximum tolerated dose}~\cite{Viossat_nature_containment}, in an effort to maximise the probability of cure.
However, aggressive treatment strategies may fail in the long term. The maximum tolerated dose eliminates cancer cells that are not resistant to treatment, leaving some treatment-resistant cancer cells to flourish in the aftermath~\cite{selection_resistance}. At this point the treatment may no longer be effective.
Alternative treatment strategies have been proposed that 
apply lower drug doses  in an optimal way~\cite{gametheory_cancer_review,Gluzman_opt_cancer} to maintain a balance between non-resistant and resistant cancer cells to prolong patient survival~\cite{optimal_control_survival,Viossat_nature_containment}.
To test and validate novel treatment strategies, a robust, reliable and accessible mathematical description of cancer development is required.

Existing mathematical models of cancer development vary in complexity.
While complex models can be more realistic, simple dynamic models make it easier to understand the critical points of cancer development and identify optimal treatment strategies. This paper presents a simple dynamic model that is process-based, captures the key mechanisms and critical points of cancer development, and reproduces a wide range of real cancer data.

For example, agent-based models (ABMs) are considered to be well suited to study evolutionary and ecological processes at different temporal and spatial scales~\cite{Franssen2019,ABM_good}. ABMs are very flexible because biological observations can be translated directly into simulation rules \cite{ABM_good_2}.
However, it is often challenging to choose sensible parameters and validate  ABMs~\cite{ABM_chall,ABM_chall_2}. 
Furthermore, the physical scale of simulations is constrained by available computational resources~\cite{scaling_issue}.
The use of  dynamic ordinary and/or partial differential equation models circumvents most of these 
drawbacks~\cite{Anderson1998,Chaplain2005}.
In the simple models, the changing number of cancer cells in a tumour is described using population  growth models, such as the logistic~\cite{carrere_optimization,LV_type} or Gompertzian 
model~\cite{Viossat_nature_containment}.
However, while describing the progression of cancer quite well, such models lack a threshold separating  growth from natural extinction, known as the Allee effect in population biology~\cite{allee_book}.
In other words, typical growth models predict that, no matter how small the initial cluster of cancer cells, a large tumour is the inevitable outcome.
This is inconsistent with  the observed processes where cancer cells, or their precursors, continuously develop and are eliminated naturally by immune suppression mechanisms~\cite{init_promo_prog_2,threshold_carcinogenesis_2}.

In this paper, we propose a novel model of cancer initiation and growth that incorporates an {\em extinction threshold}. We show that the model is consistent with a wide range of available data on cancer growth and age-specific incidence rates at the population level.
We include in the model the response of the immune system to mutations, which normally prevents cancer from developing. This naturally leads to an extinction threshold.
We also propose that the threshold need not be fixed. It can change over time, for example due to a changing immune response, resulting in a {\em moving extinction threshold}.
%
%
We then develop a simple stochastic model of cell mutations and combine it with the extinction threshold growth model. In this way, we are able to reproduce the  exponential increase in age-specific cumulative risks of most cancers using a constant immune response. 
We then go one step further and reproduce the unusual increase in the age-specific cumulative risk of breast cancer in women using  a 
time-varying immune response  
that reflects the different levels of progesterone that occur at different phases of the menstrual cycle and menopausal treatment.
In this way, we provide new insights into the dynamic mechanisms that underlie  the development of different cancers.

This paper is organised as follows.
In Sec.~\ref{sec:model}, we discuss the importance of an extinction threshold in cancer development, review the classical models of population growth with an extinction threshold, and show their limitations in accurately describing cancer development. We then propose a new dynamic model  that 
overcomes the limitations of the classical models.
In Sec.~\ref{sec:breast_cancer_data}, we demonstrate very good agreement between the proposed model and available data on untreated breast cancer progression in mice, typical age-specific cumulative risk of colorectal cancer in women and unusual
age-specific cumulative risk of breast cancer in women.
The last result is achieved by consecutively including processes related to the menstrual cycle, menopause and postmenopausal hormone replacement therapy (HRT).
\begin{figure}
    \centering
    \includegraphics[width=\textwidth]{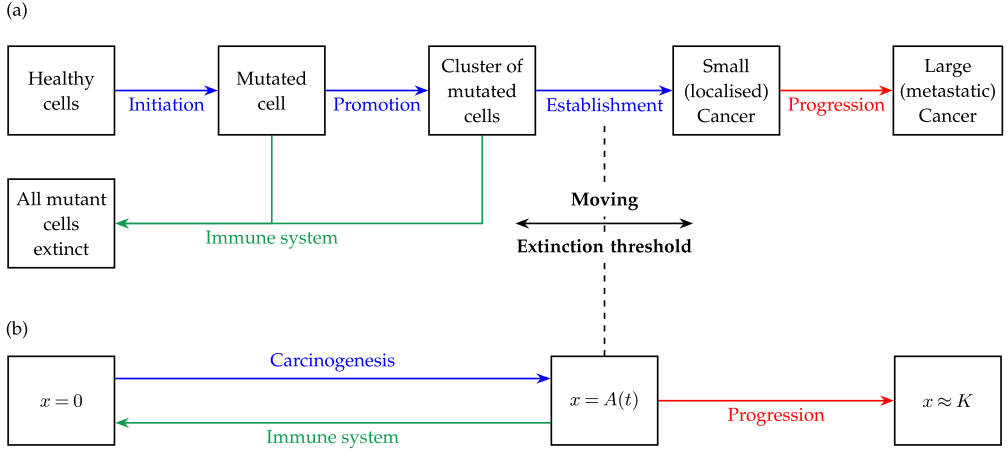}
    \caption{
Two conceptual diagrams of cancer development including (blue) carcinogenesis and (red) subsequent cancer progression. 
In (a), carcinogenesis is depicted as a three-step process. It consists of initiation, promotion and establishment, with a moving extinction threshold. In (b) carcinogenesis is simplified to a one-step process with a moving extinction threshold $A(t)$ for the sake of the simple dynamic model developed in Sec.~\ref{sec:model}. 
}
\label{fig:carcinogenesis}
\end{figure}
\noindent
Finally, new insights into  cancer development derived from the proposed model are highlighted in Sec.~\ref{sec:conclusion}.

\section{ A simple dynamic model of cancer development
}
\label{sec:model}

The aim of this section is to develop a simple dynamic model that 
\begin{itemize}
    \item 
    Captures two phases of cancer development: the onset of cancer, known as carcinogenesis, and the subsequent progression of cancer.
    \item
    Is formulated in terms of cell birth and death processes alone for clear biological interpretation.
    \item 
    Shows very good agreement with real cancer data.
\end{itemize}

We discuss the limitations of classical population growth models in capturing both phases of cancer development and propose a new model that overcomes these limitations.  One of the central concepts of our model is an {\em extinction threshold}. In the context of cancer development, this threshold is a critical size of a cluster of mutated cells below which the cluster can be eradicated by the immune system alone, and above which the cluster develops into cancer.

\subsection{Importance of an extinction threshold}
\label{sec:motivation_threshold}

Carcinogenesis can be described in terms of a three-step process consisting of {\em initiation, promotion}, and {\em establishment}, as shown in Fig.~\ref{fig:carcinogenesis}(a).

{\em Initiation} occurs when a healthy cell with the ability to divide suffers  a mutation that is beyond repair.  A {\em cell mutation} is a damage to the DNA sequence of a cell.
Mutations occur naturally as a result of oxidation, ionising radiation, UV radiation, smoking, chemicals in food and drink, and other factors.
It is estimated that, on average, a single cell can suffer up to $10^5$ instances of DNA damage per day from internal (endogenous) and external (exogenous) sources~\cite{oxidative_damage}. This raises the question of why carcinogenesis is comparatively rare and what causes only certain mutations to develop into cancer.
Typically,  genes such as tumour suppression genes or DNA mismatch repair genes 
repair damaged cells or initiate programmed death (apoptosis) of unwanted cells, 
greatly limiting the effects of mutations~\cite{p53_5000_citation,the_Tumor_Suppressor_Gene,dna_repair}. 
However, if the repair mechanisms fail,  the mutated cell undergoes {\em promotion}, which occurs when the single mutated cell rapidly divides (proliferates) to form a small cluster of mutated cells~\cite{init_promo_prog_2}. Although the mutation is irreversible at this stage, the immune system may still be able to eliminate such clusters through a variety of mechanisms~\cite{carcino_genesis_immun,immune_system_cancer_review}.
Crucially for our model, this elimination is only possible if the cluster is small enough and the immune response is strong enough~\cite{init_promo_prog_2}.
{\em Establishment} occurs when the immune system fails to eliminate the cluster, and the cluster develops into cancer.
In other words, there is an {\em extinction threshold} above which a cluster of mutated cells can no longer be eradicated by the immune system alone. 
This is evidenced, for example, by the work of Pitot and Dragan~\cite{init_promo_prog_2}  and Kirsch-Volders et al.~\cite{threshold_carcinogenesis_2}
who identify a "threshold" separating promotion from progression in cancer growth data.

In summary, without an extinction threshold, any mutated cell with the ability to divide would lead to cancer. This is clearly not the case.
Therefore, in the following subsections, we will progressively construct a simple dynamic model of cancer development that is based on the conceptual diagram in Fig.~\ref{fig:carcinogenesis}(b) and incorporates a (moving) extinction threshold.

\subsection{Cancer progression in the presence of an extinction threshold
}
\label{sec:progression_threshold}

We start with the second phase of cancer development, progression above the extinction threshold, where small cancers grow in size and potentially spread to other organs (metastasis).
We briefly review classical growth models from population biology and introduce a defining feature of cancer progression to highlight their limitations in capturing cancer progression in the presence of an extinction threshold. We then  propose a new model that avoids the limitations of classical growth models.

\subsubsection{Classical growth models with no extinction threshold}
\label{sec:Richards}

The simplest model of population growth with no extinction threshold, proposed by Thomas Malthus~\cite{malthusian_paper}, assumes unbounded {\em exponential growth}. 
A more realistic model, proposed by Pierre-François Verhulst~\cite{verhulst_paper}, accounts for resource limitation and introduces bounded {\em logistic growth} towards a {\em carrying capacity $K$} - the maximum population size that can be supported by the environment.  A specific feature of logistic growth is the {\it symmetric progression}:  growth of small populations and 
saturation of large populations as they approach carrying capacity $K$ occur at the same rate.
While the progression of certain cancers can be described by the symmetric logistic growth model~\cite{noemi_logistic,carrere_optimization}, many known cancers exhibit {\em asymmetric progression}, where  small populations grow faster than large populations saturate as they approach carrying capacity $K$~\cite{mice_data_cloud}.
These cancers are often described by the asymmetric growth model proposed by Benjamin Gompertz~\cite{Gompertz_paper}. 
In the remainder of this section, we use the asymmetry of cancer progression as its {\em defining feature}, quantified by the ratio of the growth rate of small populations to the saturation rate of large populations.

We will be interested in a more general growth model, giving rise to the so-called
{\em generalised logistic growth}, proposed by Francis John Richards~\cite{Richards_paper}. A particular advantage of Richards' model is that it can be parameterised to give different shapes of temporal growth, both symmetric and asymmetric, making it very versatile for reproducing the progression of different types of cancer.
We denote the number of cancer cells by a continuous variable $x(t)\ge 0$, and write Richards' model as an ordinary differential equation (ODE),\footnote{
The number of cancer cells is a natural number. However, for large numbers of cells, it is convenient to model them using real numbers.}
\begin{equation}
   \frac{dx}{dt} = \nu rx - \nu \mu x^{\frac{\nu+1}{\nu}} = \nu r x \left(1 - \left(\frac{x}{K} \right)^{\frac{1}{\nu}} \right),
    \label{eq:richard}
\end{equation}
where $t$ is time, $dx/dt$ is the instantaneous rate of change of the number of cancer cells, $\nu r > 0$ is a constant per capita growth rate, $\nu \mu x^{1/\nu} \ge 0$ is a population-dependent per capita death rate  that increases with $x$, $\nu > 0$ is the shape parameter and $K= (r/\mu)^\nu$ is the carrying capacity.\footnote{A per capita rate of change is the rate of change for each cell, or per unit of population, and is given by $(dx/dt)/x$.}
We now describe the properties of the Richards' model and refer to~\cite[Sec. 1.1]{sup} 
for technical details.

The Richards'  model has two stationary (fixed in time) solutions, also called equilibrium points,
$$
 x = 0\quad\mbox{and}\quad x=K.
$$
Linear stability analysis reveals that the extinction $x = 0$ is exponentially unstable with  divergence rate $\lambda_{grow} = \nu r$, while the carrying capacity $x = K$ is exponentially stable with convergence (saturation) rate 
$-\lambda_{sat} = r$.
Thus,  the defining feature of cancer progression in this model is
$$
\left|\frac{\lambda_{grow}}{\lambda_{sat}}\right| =\nu.
$$
In other words, small populations $x(t)$ in equation~\eqref{eq:richard} grow exponentially at the rate $\nu r$, while large populations saturate exponentially at the rate $r$ as they approach $K$. Thus, setting $\nu =1$ gives the Verhulst model with symmetric logistic growth, and setting $\nu > 1$ gives a model with asymmetric growth, where small populations grow faster than large populations  saturate at $K$.
The Gompertz model with asymmetric Gompertzian growth is obtained in the limit $\nu\to \infty$. 
We note that small populations grow more slowly than large populations saturate  at $K$ when $0 < \nu < 1$, and to the best of our knowledge there are no examples of such cancer progression. 
Therefore, in order to reproduce typical cancer progression, it is necessary that the defining feature of cancer progression satisfies
\begin{equation}
    \left|\frac{\lambda_{grow}}{\lambda_{sat}}\right| \ge 1.
    \label{eq:ratio_desired}
\end{equation}
This requirement is represented in Fig.~\ref{fig:eigenvalue_ratio_v2} by the blue 
region. The Richards model~\eqref{eq:richard}, represented by the black curve in Fig.~\ref{fig:eigenvalue_ratio_v2},  falls within the desired blue region when $\nu>1$.

\subsubsection{Model 1: A classical growth  model with an extinction threshold}
\label{sec:volterra_section}

The simplest model of  population growth with an extinction threshold was proposed by Vito Volterra~\cite{Volterra}, who modified  Verhulst's model.
In our first attempt to introduce an extinction threshold, we modify Richards' model~\eqref{eq:richard} in a similar way.
We call this Model 1,
\begin{figure}[]
    \centering
    \includegraphics[width=0.6\textwidth]{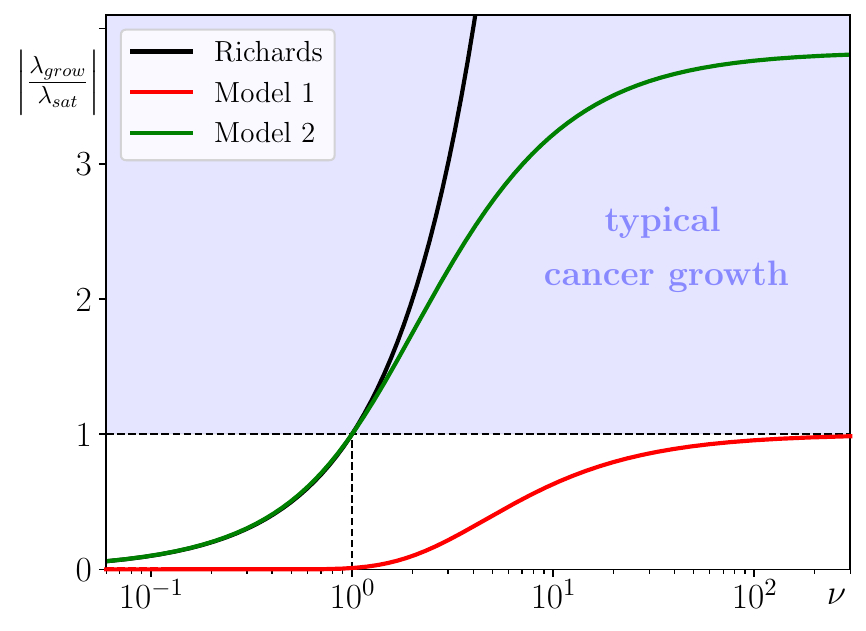}
    \caption{
    The defining feature of cancer progression
    (the ratio of the small cancer's growth rate $\lambda_{grow}$ to the large cancer's saturation rate $\lambda_{sat}$)  
    vs. the growth parameter $\nu$ for: (black) the Richards' model~\eqref{eq:richard} with no extinction threshold, (red) 
    classical Model 1 in Eq.~\eqref{eq:gen_Allee} with an extinction threshold, and (green) proposed Model 2 in Eq.~\eqref{eq:richard_harvest} with an extinction threshold. $r=1$, $A=1$, and  $K=100$. (In Model 2, $\mu$ and $s$ have been adjusted for each value of $\nu$ to keep $A=1$ and  $K=100$). Note the logarithmic scale for $\nu$.
    }
    \label{fig:eigenvalue_ratio_v2}
\end{figure}
\begin{equation}
    \begin{split}
        \frac{dx}{dt}
        &=-\nu^2 r\, x  + 
        \nu^2 \mu\, x^\frac{\nu + 1}{\nu} -
        \nu^2 \gamma\, x^\frac{\nu + 2}{\nu}\\
        &=-\nu^2 r\, x 
        \left(1-\left(\frac{x}{A}\right)^{\frac{1}{\nu}}\right)
        \left(1-\left(\frac{x}{K}\right)^{\frac{1}{\nu}}\right),
    \end{split}
    \label{eq:gen_Allee}
\end{equation}
where $t$ is time, $dx/dt$ is the instantaneous rate of change of the number of cancer cells,
$\nu^2 r > 0$ becomes a constant per capita death rate of small populations, $\nu^2 \mu x^{1/\nu} \ge 0$ becomes a population-dependent per capita birth rate that increases with $x$ and $\nu > 0$ is the shape parameter. The third term is new and corresponds to a population-dependent per capita death rate of large populations $\nu^2 \gamma  x^{2/\nu} \ge 0$, which also increases with $x$. We now describe the properties of Model 1 and refer to ~\cite[Sec.1.2]{sup} for technical details.

For a suitable choice of the new parameter $\gamma$,  which quantifies the death rate of large populations, Model 1 has three equilibrium points,
$$
 x = 0,\quad  x=A\quad\mbox{and}\quad x=K,
$$
including an extinction 
threshold  $0 < A < K$, given by
$$
A = \left(\frac{\mu - \sqrt{\mu^2 - 4 r\gamma}}{2\gamma}\right)^{\nu}.
$$
The most commonly used version of Model 1, obtained by setting $\nu=1$ in Eq.~\eqref{eq:gen_Allee}, is the original Volterra model~\cite{Volterra}. 
On the other hand, in the limit $\nu\to \infty$, we recover 
a less commonly used version of Model 1, namely the Gompertz model with an extinction threshold~\cite{gompertz_allee}.

\begin{figure}[H]
    \centering
    \includegraphics[width=\textwidth]{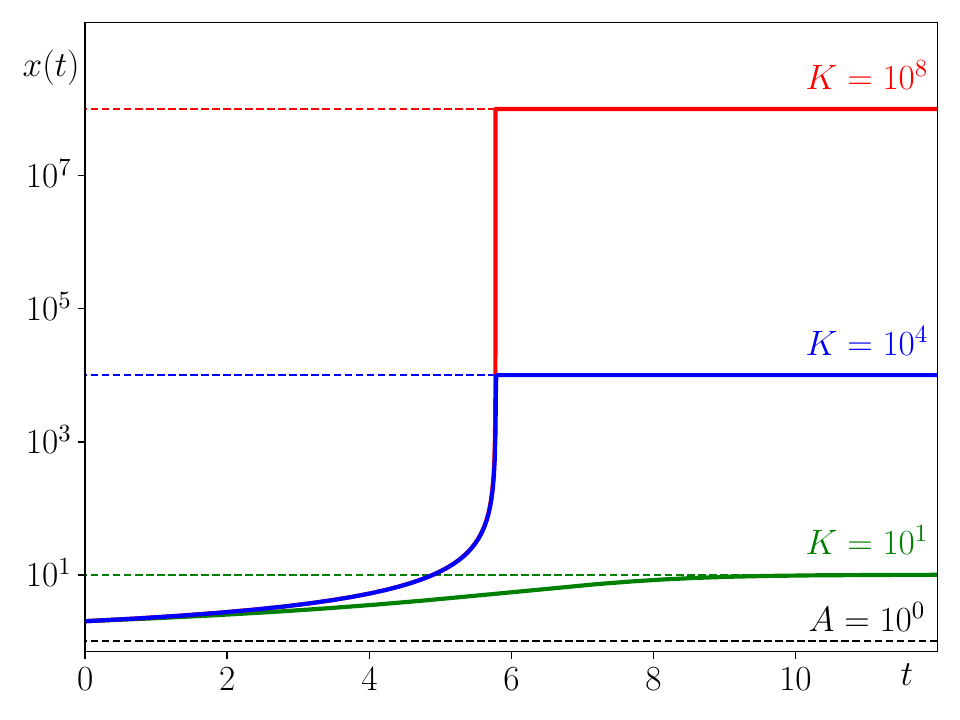}
    \caption{
    Population growth $x(t)$ in the commonly used version of Model 1 (Eq.~\eqref{eq:gen_Allee} with $\nu=1$) started from $x(0)= 2$ with a per capita decay rate $r= 0.12$, extinction threshold $A=1$ and three different values of carrying capacity: (green) $K=10$ , (blue) $K=10^4$ and (red) $K=10^8$.  For realistic values of $K$, the solutions show an unusual step-like  behaviour that does not resemble the progression of any known cancer. For $K=10^4$, the (blue) progression curve takes about $5.5$ time units to reach $1\%$ of $K$, and only $10^{-1}$ time units to grow from $1\%$ to $99\%$ of $K$. For $K=10^8$, the (red) progression curve takes a similar time to reach $1\%$ of $K$ and only $10^{-5}$ time units to grow from $1\%$ to $99\%$ of $K$. 
    Such population jumps are also unusual in ecology~\cite{different_AlleeVersion}.
    Note the logarithmic scale for $x(t)$.
    }
    \label{fig:proposition_illustration}
\end{figure}

Linear stability analysis reveals that 
the extinction threshold $x=A$ is exponentially unstable with divergence rate $\lambda_{grow}$, and the carrying capacity $x=K$ is exponentially stable with convergence (saturation) rate $-\lambda_{sat}$.
However, the introduction of an extinction threshold by this approach has major limitations in reproducing typical cancer progression. 
To formulate these limitations rigorously, we prove in~\cite[Sec.1.2]{sup} 
that, in Model 1 with extinction threshold $A$,
small populations always grow more slowly than large populations saturate as they approach $K$. 
In other words, the defining feature of cancer progression in Model 1,
\begin{equation}
    \left|\frac{\lambda_{grow}}{\lambda_{sat}}\right| = \left(\frac{A}{K}\right)^\frac{1}{\nu} < 1\quad\mbox{for all}\quad \nu>0,
    \label{eq:ratio_nu}
\end{equation}
is in disagreement with the progression of known cancers.
This is further illustrated in Fig.~\ref{fig:eigenvalue_ratio_v2}, where the 
defining feature $|\lambda_{grow}/\lambda_{sat}|$ of cancer progression in Model 1, represented by the red curve, does not overlap with the desired blue region for any $\nu>0$.
Moreover, in typical cancer development, $A$ is several orders of magnitude smaller than $K$. Thus, if $\nu=1$ (the most commonly used  Volterra model),
small populations grow several orders of magnitude slower than large populations saturate  at $K$; the red curve in Fig.~\ref{fig:eigenvalue_ratio_v2} is nowhere near the desired blue region when $\nu=1$. 
This has serious implications for the shape of the temporal cancer progression curve. Namely, populations stay just above $A$ for a long time before approaching $K$ sharply, in a somewhat unrealistic jump similar to a step function; this is illustrated and quantified in Fig.~\ref{fig:proposition_illustration}.
At best, when $\nu\to\infty$ (the less commonly used Gompertz model with an extinction threshold), the defining feature $|\lambda_{grow}/\lambda_{sat}|$ approaches 1 from below and progression becomes symmetric; the red curve in Fig.~\ref{fig:eigenvalue_ratio_v2} approaches the desired blue region from below for large $\nu$.
 
In summary, a different approach to an extinction threshold is required to achieve the desired asymmetry~\eqref{eq:ratio_desired} that is characteristic of typical cancer progression.

\subsubsection{Model 2: A new growth model with an extinction threshold}
\label{sec:newmodel}

In this section, we construct a new model of cancer progression that builds on the strengths of the classical growth models and overcomes their limitations. We aim to retain:
\begin{itemize}
    \item
    An extinction threshold.
    \item 
    The versatility of growth above the extinction threshold, ranging from symmetric to the desired asymmetric progression~\eqref{eq:ratio_desired}.
    \item 
    Formulation in terms of birth and death processes alone for clear biological interpretation.  
\end{itemize}

Guided by the results of the previous section and the bullet points above, we start with the Richards' model~\eqref{eq:richard} and introduce a new process that accounts for cancer suppression mechanisms. In this process, the immune system prevents or eliminates a certain  number of mutated cells per day.\footnote{This process is similar to harvesting in population biology.} 
In the model, we represent this process by an additional cell death term $-s$.
Furthermore, we recognise that the immune response will vary in time $t$, possibly in a complicated way  and on different time scales, e.g.  due to seasonal variations, viral infections, the menstrual cycle, immune deficiencies, immunotherapy, ageing etc. We therefore allow the additional term $-s(t)$ to vary over time. This means that the immune system prevents or eliminates a certain number of mutated cells each day, and this number changes over time as different factors influence the strength of the immune system.
This gives us Model 2:
\begin{equation}
    \frac{dx}{dt} =
    \begin{cases}
        \nu r x - \nu \mu x^{\frac{\nu+1}{\nu}} - s(t)& \text{ if }x>0,\\
        0&\text{ if }x=0,
    \end{cases}
    \label{eq:richard_harvest}
\end{equation}
where $t$ is time, $dx/dt$ is the instantaneous rate of change of the number of cancer cells, $\nu r > 0$ is a constant per capita growth rate, $\nu \mu x^{1/\nu} \ge 0$ is a population-dependent per capita death rate  that increases with $x$ and $\nu > 0$ is the shape parameter.
We now describe the properties of Model 2 and refer to~\cite[Sec. 1.3]{sup} for technical details.

Owing to the time-varying $s(t)$, the non-autonomous Model 2 has only one equilibrium, namely extinction $x=0$. This equilibrium is always stable and, in contrast to Model 1, the population $x(t)$ becomes extinct in finite time.

To gain further insight into the behaviour of Model 2, it is useful to start with a simplified description, where the immune response 
is constant (fixed in time), that is $s(t) = s = \mbox{const.}$
Then, for a suitable choice of $s$, Model 2 has three equilibrium points,
\begin{equation}
    x=0,\quad x = A \quad\mbox{and} \quad x =  K,
    \label{eq:eq_Model2}
\end{equation}
where $0< A < K$.
While there is no closed-form formula for $A$, the position of the {\em stationary extinction threshold} $A\sim 10^a$ can be approximated by
\begin{equation}
\label{eq:Amodel2}
A \approx \frac{s}{\nu r - \mu(\nu + a\,\ln{10})}.
\end{equation}
This approximation allows us to identify what we call {\em threshold parameters}: the  immune response $s$, the per capita growth rate $\nu r$ and the death rate parameter $\nu\mu$. It also shows how the position of the extinction threshold changes with different threshold parameters.

Linear stability analysis reveals that the stationary extinction threshold $x=A$ is exponentially unstable with divergence rate $\lambda_{grow}$, and the carrying capacity $x=K$ is exponentially stable with convergence (saturation) rate $-\lambda_{sat}$. Most importantly, we prove in~\cite[Sec.1.3]{sup}
that small populations grow faster than large populations saturate as they approach $K$ if the growth parameter $\nu$ is set greater than one. To be specific,  the defining feature of cancer progression in Model 2,
\begin{equation*}
    \left|\frac{\lambda_{grow}}{\lambda_{sat}}\right| 
    \geq 1\quad\mbox{for all}\quad \nu\geq1,
\end{equation*}
is in agreement with the progression of known cancers.
This is further illustrated in Fig.~\ref{fig:eigenvalue_ratio_v2}, where the defining feature $|\lambda_{grow}/\lambda_{sat}|$ of cancer progression in Model 2 with stationary extinction threshold, represented by the green curve, is in the desired blue region for $\nu \ge 1$.

In the more realistic description, some threshold parameters  will vary over time. This means that the threshold will also vary over time and one will have to consider a {\em moving extinction threshold} $A(t)$. Our focus will be on a moving extinction threshold due to a time varying immune response $s(t)$.


In summary, Model 2 incorporates an extinction threshold in a way that is process-based, preserves the versatility of different growth functions and, most importantly, retains the progression characteristics of typical cancers.

\subsection{Carcinogenesis: random cell mutations in the presence of an extinction threshold}
\label{sec:carcinogenesis}

We now consider the first phase of cancer progression, carcinogenesis, during which the tumour is initiated. Crucially, we model carcinogenesis as an interplay between random cell mutations and the extinction threshold. We consider a fixed number of healthy stem cells $n$. 
We then assume that there is a small probability $p(t)$ that each healthy stem cell will mutate in {\em one day}~\cite{1_day_devision}.
\begin{figure}[]
    \centering
    \includegraphics[width=1.\textwidth]{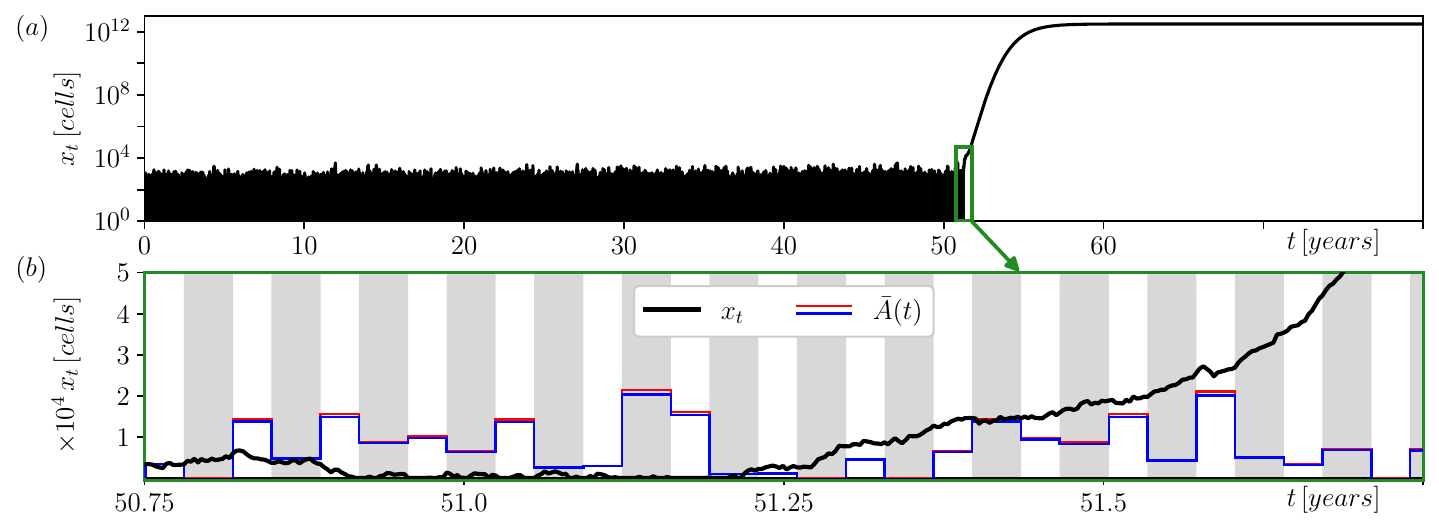}
    \caption{
    (a) An example of breast cancer development $x(t)$ in model~\eqref{eq:stochastic_mapping_moving_A} for a single realisation of random mutations $m(t)$ using the parameter values in~\cite[Table 3]{sup},
    and for periodic variations in the immune response $s_m(t)$ due to menstrual cycle in Eq.~\eqref{eq:moving_A_def_micro}.
    (b) An extended view of the critical point around age 51.25, when the (black) number of mutated cells $x(t)$ exceeds the (blue/red) moving extinction threshold $\bar{A}(t)$
    and develops into cancer.
    $s_m(t)$ takes values: (grey stripes) $s_{min}= np_0 + 262$  cells/day 
    during  the 14-day long luteal phases with a weaker immune response, and (white stripes) $s_{max}= np_0 +354$ cells/day during  the 11-day long follicular phases with a stronger immune response; see Fig.~\ref{fig:movingA_global} ahead for more details.
    In each phase, the time average $\overline{m}$ of $m(t)$ is used to obtain $\bar{A}(t)$ (red) numerically and (blue) using the approximation in Eq.~\eqref{eq:approx_stoch_A} and
    the parameter values in~\cite[Table 3]{sup}.
    Note the logarithmic scale for $x_t$ in (a).
    }
    \label{fig:example_traj}
\end{figure}
\noindent
In other words, the mutation of a single cell is a {\em Bernoulli trial}~\cite{bernoiulli} with a small probability of success (mutation) $p(t)$. If we additionally assume that the mutations of individual cells are independent events, the mutation of a group of $n$ cells is a {\em binomial process}~\cite{binomial_process_textbook}.
In such a process, the probability $P$ that $m$ of $n$ healthy cells will mutate in one day is given by the binomial distribution.
\begin{equation}
    P\left(m;n,p(t)\right)
    = 
    \frac{n!}{m!(n - m!)}\,\,p(t)^m\,\left(1-p(t)\right)^{(n - m)},
    \label{eq:binomial_dist_1delta}
\end{equation}
with expected value $np(t)$ and variance $np(t)(1-p(t))$.

An important aspect of our model is that the probability $p(t)$ that a single cell will mutate in one day evolves slowly over a lifetime. We assume that $p(t)$ starts from $p(0)=p_0$ at birth, increases linearly with time $t$,
\begin{equation}
    p(t) = p_0\left(1 + \frac{\delta\,t}{100\,T}\right),
    \label{eq:p_t}
\end{equation}
and reaches a $\delta$ percent increase at life expectancy $T$.
This means that the binomial distribution in~\eqref{eq:binomial_dist_1delta} also evolves over a lifetime, as shown in~\cite[Fig.1]{sup}.

Our model of carcinogenesis can be understood in terms of two competing processes. On the one hand, random cell mutations lead to clusters of mutated cells. On the other hand, the immune system tries to eradicate these clusters, and its strength is represented by the position of the extinction threshold. Meanwhile, the mutation rate increases slowly with age, and the extinction threshold moves over time to reflect, for example, changes in the strength of the immune system. 
Whether a cluster of mutated cells can be eliminated by the immune system alone or develops into a small cancer depends on its size relative to (i.e. below or above) the current position  of the extinction threshold. 

\subsection{The complete cancer development model}

In this section we  combine the cancer progression Model 2 proposed in  Sec.~\ref{sec:model}~\ref{sec:progression_threshold}~\ref{sec:newmodel} and the random mutation model proposed in  Sec.~\ref{sec:model}~\ref{sec:carcinogenesis} to construct a {\em complete model} of cancer development.

In the first step, we fix a short time interval $\Delta$ during which the size of the cluster and the immune response do not change significantly. We then integrate Eq.~\eqref{eq:richard_harvest} from $t$ to $(t+\Delta)$ and obtain the {\em development function} for the time interval $\Delta$,
$$
F_\Delta(x,t) = x(t) + \Delta\left[\nu r x(t) - \nu \mu x(t)^{\frac{\nu+1}{\nu}} - s_m(t)\right] + m_\Delta(t).
$$
Here,  $x(t)$ is the current cluster or cancer size. The second term is its increase or decrease during the subsequent time interval $\Delta$, which results from the combination of intrinsic growth and elimination of cancer cells by the immune system $s_m(t)$.\footnote{The immune response in the presence of mutations is given by $s_m(t) = s(t) + \langle m_\Delta\rangle(t)$, where $s(t)$ is the immune response in the absence of mutations and  $\langle m_\Delta\rangle(t)$ is the slowly-varying expected value of new mutations $m_\Delta(t)$.}
The last term, $m_\Delta(t)$, represents new mutations that occur during the time interval $\Delta$. 

Next, we fix $\Delta = 1$ day, and consider the daily development function
$$
F(x,t) = x(t) + \left[ \nu r x(t) - \nu \mu x(t)^{\frac{\nu+1}{\nu}} - s_m(t)\right] + m(t),
$$
\begin{figure*}[ht]
    \centering
    \includegraphics[width=\textwidth]{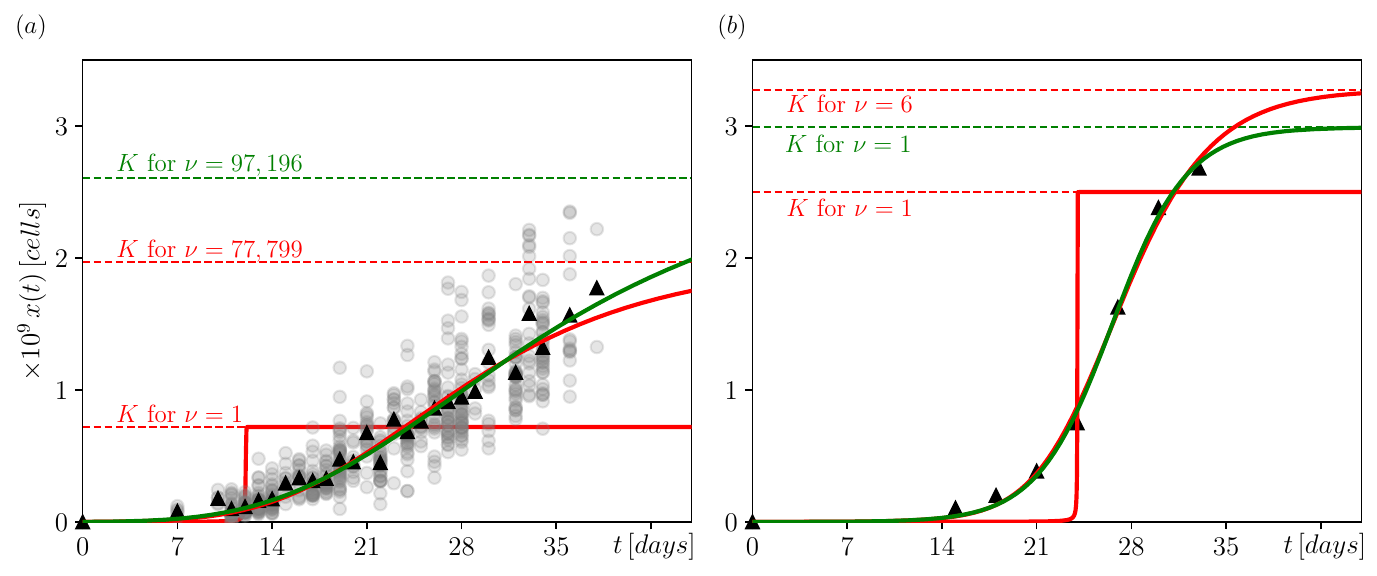}
    \caption{ 
    A comparison between real data on breast cancer progression in mice and dynamic models of cancer progression.
    (a) Dataset 1 from~\cite{mice_data_cloud}  shown as (circles) individual measurements of tumour size and (triangles) mean tumour size for each day, together with the best theoretical fits using (red curves) Model 1 and (green curve) Model 2. 
    (b) Dataset 2 from~\cite{treatment_mice} shown as (triangles) mean tumour size for each day, together with the best theoretical fits using (red curves) Model 1 and (green curve) Model 2.
    The model solutions start from the first data point at time $t=0$, the parameter values are listed in~\cite[Table 1]{sup}, and the resulting carrying capacities are shown as dashed horizontal lines.
    }
    \label{fig:fit_rh_volt_fix_x0}
\end{figure*}
\noindent
where $m(t)$ represents new mutations in one day
obtained from the probability distribution in Eq.~\eqref{eq:binomial_dist_1delta}.  

In the final step, we write our cancer development model in terms of the daily development function $F(x,t)$, as a difference equation with a time step interval of 1 day,
\begin{align}
    x(t+1)=
    \begin{cases}
        F(x,t)  &\text{if }\quad F(x,t) \ge 0,\\
        0 &\text{if } \quad F(x,t) <0.
    \end{cases}
    \label{eq:stochastic_mapping_moving_A}
\end{align}
The instantaneous position of the {\em moving extinction threshold} $A(t)\sim 10^a$ in model~\eqref{eq:stochastic_mapping_moving_A} can be approximated by
\begin{equation}
    \bar{A}(t) \approx \frac{s_m(t) - \overline{m}(t)}{\nu r-\mu(\nu + a\,\ln{10})},
    \label{eq:approx_stoch_A}
\end{equation}
where $\overline{m}(t)$ is the time average of $m(t)$ over a chosen time interval during which $s_m(t)$ does not change 
significantly; see~\cite[Sec. 1.5]{sup} for more details. 
This approximation shows that  the extinction threshold for the complete model can move over time due to  the randomly varying mutations, $\overline{m}(t)$,  or a combination of  the changing immune response, $s_m(t)$, and $\overline{m}(t)$.

An interesting consequence of a moving extinction threshold is that if the cluster of mutated cells $x(t)$ exceeds the moving threshold $\bar{A}(t)$, there are two possible scenarios. Often, the cluster of mutated cells remains above the moving threshold and progresses directly into cancer. However, there is another possibility, known as the \enquote{rescue event}~\cite{hassan_stoch_res}: the threshold rises faster than the growth of the cluster of mutated cells, the cluster is soon back below the moving threshold, and the transition to cancer is avoided; see~\cite[Sec.1.5]{sup} for more details.

Figure~\ref{fig:example_traj}(a) shows 
an example of cancer development $x(t)$ in model~\eqref{eq:stochastic_mapping_moving_A} for a single realisation of random mutations $m(t)$ and periodically varying $s_m(t)$ due to the menstrual cycle.
Figure~\ref{fig:example_traj}(b) shows the moment when the cluster of mutated cells crosses the moving extinction threshold $\bar{A}(t)$  around the age of 51.25 and progresses into cancer.  Noteworthy is the rescue event about 5 months earlier, around the age of 50.85. 

\section{Breast cancer case study}
\label{sec:breast_cancer_data}

In this section, we use our cancer development model to reproduce real cancer data and give new insights into breast cancer development in women. 
This discussion should be seen as a proof of concept of how simple dynamic models with an extinction threshold can be used to provide a qualitative description of cancer development and identify the underlying mechanisms and critical points.

\subsection{Breast cancer progression in mice: A comparison between Models and data}
\label{sec:comparison1}

Data on the progression of untreated breast cancer in women is unavailable, hence we examine the ability of Model 1 and Model 2 to reproduce two different  datasets on the progression of untreated breast cancer  in mice. 
To do this, we combine a shooting method to solve each model and a curve fitting algorithm to estimate the model parameters~\cite[Sec. 2.1]{sup}. 
We show that Model 2 can closely reproduce both datasets and give realistic estimates of the carrying capacity in both datasets.\footnote{ In this section, in Model 2, we use a constant immune response $s(t)= s ={\it const.}$ to analyse progression above the threshold.}
In contrast, Model 1 fails to reproduce dataset 1.


Dataset 1 from Vaghi {\em et al.}~\cite{mice_data_cloud} consists of 583 data points from 65 Severe Combined Immunodeficient (SCID) mice, collected over 38 days (gray dots in Fig.~\ref{fig:fit_rh_volt_fix_x0}(a)). SCID mice exhibit severe immunodeficiency because of a lack of functional T and B  lymphocytes~\cite{Bosma_1983_527}. However, they retain natural killer cells and preserve some anti-tumour activity, albeit at a much lower level than normal~\cite{DEWAN_2005_S375}. 
Each data point represents the tumour volume in one mouse at a point in time. Just for comparison with dataset 2, we also plot the average tumour volume at a point in time (black triangles in Fig.~\ref{fig:fit_rh_volt_fix_x0}(a)).
Dataset 2 from Cabeza {\em et al.}~\cite{treatment_mice} consists of 8 data points from 10 immunocompetent C57BL/6 mice, collected over 33 days (black triangles in Fig.~\ref{fig:fit_rh_volt_fix_x0}(a)). Here, each data point represents the average tumour volume at a point in time.

Both datasets were obtained by injecting $x_0$ breast cancer cells into mice and monitoring the subsequent tumour volume progression without treatment until the mice died.
To fit the models to the data, we convert the tumour volume to the number of cancer cells, assuming a cancer cell density of $10^9$ cells/cm$^3$~\cite{1cm3_cell_count}.
The model parameters that give the best fit are listed in~\cite[Table 1]{sup}.

We begin with Model 1 in its factorised form~\eqref{eq:gen_Allee}.
First, we set $\nu=1$ to get the commonly used Volterra model and let the curve fitting algorithm estimate $r, A$ and $K$.
The best fit gives (red) step-like progression curves in Figs.~\ref{fig:fit_rh_volt_fix_x0}(a) and~\ref{fig:fit_rh_volt_fix_x0}(b), which do not reproduce any of the properties of datasets 1 and 2, respectively.
Secondly, we release $\nu$ and let the curve fitting algorithm estimate $\nu$ along with $r, A$ and $K$. 
For dataset 1, the best fit gives $\nu\approx77,800$ and a smoother (red) progression curve in Fig.~\ref{fig:fit_rh_volt_fix_x0}(a) that only fails to reproduce the later stage of cancer progression. In particular, Model 1 underestimates the carrying capacity and returns a value of $K\approx 1.967\times 10^9$ cancer cells that is below some data points. 
This is inconsistent with the fact that a patient dies some time before the tumour reaches its carrying capacity~\cite{Norton_Breast_K}.
For dataset 2, the best fit gives $\nu=6$ and a smoother (red) progression curve in Fig.~\ref{fig:fit_rh_volt_fix_x0}(b) that closely reproduces the cancer progression. 

Next, we use Model 2 in~\eqref{eq:richard_harvest} and let the curve fitting algorithm estimate its four parameters $\nu, r, \mu$ and $s$. 
The best fit gives (green) progression curves in Figs.~\ref{fig:fit_rh_volt_fix_x0}(a) and~\ref{fig:fit_rh_volt_fix_x0}(b), which are in very good agreement with datasets 1 and 2, respectively. In particular, both curves give realistic carrying capacities $K$.\footnote{Unlike Model 1, Model 2 has no prescribed $A$ and $K$, and their values must be obtained 
as equilibrium points for Model 2.} In Fig.~\ref{fig:fit_rh_volt_fix_x0}(a) for dataset 1, the large value of $\nu=97,196$ indicates an asymmetric (practically Gompertzian) progression, which explains why dataset 1 cannot be fitted by Model 1. 
In Fig.~\ref{fig:fit_rh_volt_fix_x0}(b) for dataset 2, the value of $\nu=1$ indicates symmetric (logistic) progression, which explains why this dataset can also be fitted by Model 1 with larger $\nu$.

\begin{figure}[]
    \centering
    \includegraphics[width=0.5\textwidth]{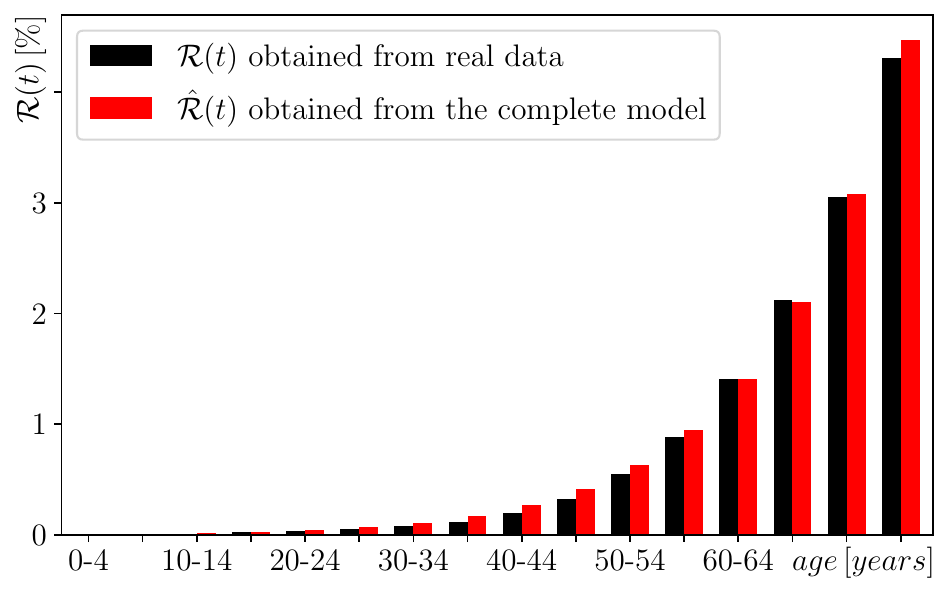}
    \caption{
    The age-specific cumulative risk of a woman developing colorectal cancer by a given age $t$. A comparison between the  (black) actual risk $\mathcal{R}(t)$ obtained from real data in~\cite{NCRI} and the (red) risk $\hat{\mathcal{R}}(t)$ obtained from the complete model~\eqref{eq:stochastic_mapping_moving_A} with constant immune response $s(t) = s_{max} = const.$
    We used the parameter values given in~\cite[Table 2 and 3]{sup} and refer to~\cite[Sec. 2.2.4]{sup} for more details.
    }
    \label{fig:crc_risk}
\end{figure}

Since the two groups of mice have significantly different immune responses, it is natural to ask whether this is captured by
Model 2 as different extinction thresholds $A$. 
Comparing the results of the model for the two groups of mice, it can be seen that the predicted extinction threshold $A\approx2\times 10^2$ cells for the immunodeficient SCID mice (dataset 1) is indeed significantly lower compared to $A\approx3\times 10^5$ cells for the immunocompetent C57BL/6 mice (dataset 2); see~\cite[Table 1]{sup}.

\subsection{Breast carcinogenesis in women: Reproducing unusual cumulative risk}
\label{sec:comparison2}

The {\em cumulative risk of a particular cancer at age $t$} is the probability, expressed as a percentage, that a person will develop that cancer by age $t$~\cite{cumulative_risk_def}.
The {\em age-specific cumulative risk} is the cumulative risk as a function of age $t$.
The ability of a mathematical model to reproduce  the cumulative risk of cancer depends crucially on its ability to capture the key mechanism(s) of carcinogenesis.
Here, we use our complete cancer development model~\eqref{eq:stochastic_mapping_moving_A} to reproduce qualitatively different age-specific cumulative risks of different cancers.


The actual age-specific cumulative risk,  which we denote by $\mathcal{R}(t)$, can be estimated from {\em age-specific incidence rate data}: the ratio of 
people diagnosed with a particular cancer to the number of all people in different age groups.
To highlight different mechanisms of cancer development, we use the age-specific incidence rates of colorectal cancer and breast cancer in women in Ireland. 
Data were obtained from the National Cancer Registry Ireland~\cite{NCRI}, for the years 1994-2021 for colorectal cancer~\footnote{
In \cite{NCRI}, select ``By age" in the header, ``Colorectal|C18-C21" under Cancer, ``Ireland" under Region, ``Females" under Sex, ``1994" under Year from, and ``2021" under Year to.} and for a census year 2016 for breast cancer~\footnote{
In \cite{NCRI}, select ``By age" in the header, ``Breast|C50" under Cancer, ``Ireland" under Region, ``Females" under Sex, ``2016" under Year from, and ``2016" under Year to.}.
\begin{figure*}
    \centering
    \includegraphics[width=\textwidth]{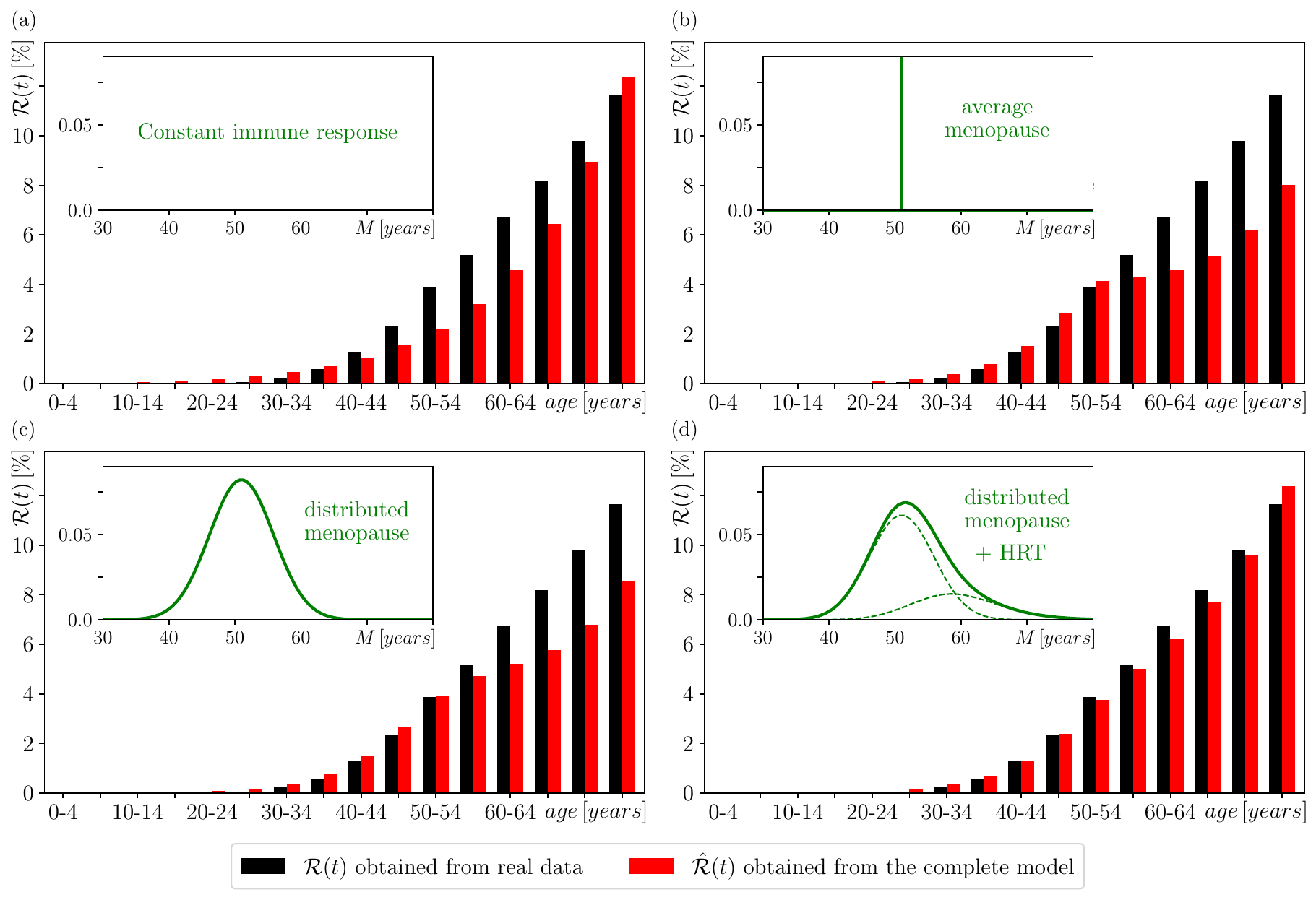}
    \caption{
     The  age-specific cumulative risk of a woman developing breast cancer by a given age $t$. A comparison between the  (black) actual risk $\mathcal{R}(t)$ obtained from real data in~\cite{NCRI} and the (red) risk $\hat{\mathcal{R}}(t)$ obtained from the complete model~\eqref{eq:stochastic_mapping_moving_A} with four different forms of immune response $s(t)$.
     (a) A constant immune response with $s(t) = s_{max} = const$ gives the (red) risk $\hat{\mathcal{R}}(t)$, which does not match the (black) actual risk $\mathcal{R}(t)$.
    (b) A periodically varying immune response~\eqref{eq:moving_A_def_micro} between different levels during different phases of the menstrual cycle, with a fixed age of menopause $M = 51$ years, gives good agreement with the (black) actual risk $\mathcal{R}(t)$ only up to the age band 50-54.
    (c) A periodically varying immune response~\eqref{eq:moving_A_def_micro} with
    normally distributed $M$ gives a better agreement with the (black) actual risk $\mathcal{R}(t)$ but still underpredicts $\mathcal{R}(t)$ within the age bands 50-75.
    (d) A periodically varying immune response~\eqref{eq:moving_A_def_micro} with $M$ that is
    normally distributed and then delayed for 26\% of women due to hormone replacement therapy (HRT) closely reproduces the (black) actual risk $\mathcal{R}(t)$. We used the parameter values in~\cite[Tables 2 and 3]{sup}.
    }
    \label{fig:breast_risk}
\end{figure*}
\noindent
The data consist of incidence rates for 16 age groups of 5 years each, which we have used to calculate the $\mathcal{R}(t)$ shown in black for colorectal cancer in Fig.~\ref{fig:crc_risk}  and breast cancer in Fig.~\ref{fig:breast_risk};
see~\cite{cum_risk_explanation} and~\cite[Sec. 2]{sup} for more details.

The actual age-specific cumulative risk of colorectal cancer in women, (black) $\mathcal{R}(t)$ in Fig.~\ref{fig:crc_risk},  appears to increase exponentially with age~\cite[Fig.~4]{sup}. 
However, this is not the case for breast cancer in women,  (black) $\mathcal{R}(t)$ in Fig.~\ref{fig:breast_risk}. The striking feature of the actual  age-specific cumulative risk of breast cancer  in women is a (7th order) polynomial increase up to the age of 44-50, after which it changes and becomes linear~\cite[Fig.5]{sup}.
This is in contrast to the vast majority of cancers, which appear to show an exponential increase in age-specific cumulative risk across all age groups; for example, see~\cite[Fig.~6]{sup} for blood cancer and brain cancer in women.

We denote the {\em theoretical  age-specific cumulative risk} by $\hat{\mathcal{R}}(t)$. $\hat{\mathcal{R}}(t)$ is obtained by a combination of Monte Carlo simulations of our model~\eqref{eq:stochastic_mapping_moving_A} and a fitting algorithm, both of which are described in~\cite[Sec.3]{sup}.
We recall that the model is based on the hypothesis that carcinogenesis results from two competing processes: the random formation of clusters of mutated cells at a rate that increases with age, and the elimination of mutated cells by the immune response that can also  vary in time.
To gain new insights into the  different mechanisms that underlie the development of different types of cancer, we consider two cases:
(i)  a fixed immune response, and
(ii) a time-varying immune response.
The model parameters that give the best fits $\hat{\mathcal{R}}(t)$ to the actual age-specific cumulative risks $\mathcal{R}(t)$ are listed in~\cite[Table 2]{sup}.

\subsubsection{Typical  age-specific cumulative risk of most cancers: Fixed immune response}

We start with a constant immune response, 
$$
s_m(t) = s_{max} = const.,
$$ 
which gives rise to a randomly moving extinction threshold~\eqref{eq:approx_stoch_A} due to mutations $m(t)$ alone.

The fixed immune response model predicts an age-specific cumulative risk $\hat{\mathcal{R}}(t)$ that increases exponentially with age.
This exponential increase is in excellent agreement with the (black) actual risk $\mathcal{R}(t)$ for colorectal cancer in women in Fig.~\ref{fig:crc_risk}, and appears to be consistent with the age-specific cumulative risk for many other cancers~\cite[Fig.~6]{sup}.
%
However, this $\hat{\mathcal{R}}(t)$ does not capture the unusual  age-specific cumulative risk of breast cancer in women. In Fig.~\ref{fig:breast_risk}(a), the (red) $\hat{\mathcal{R}}(t)$ overestimates the (black) $\mathcal{R}(t)$ below the age of 35 and underestimates it above the age of 35. 
\begin{figure*}[]
    \centering
    \includegraphics[width=\textwidth]{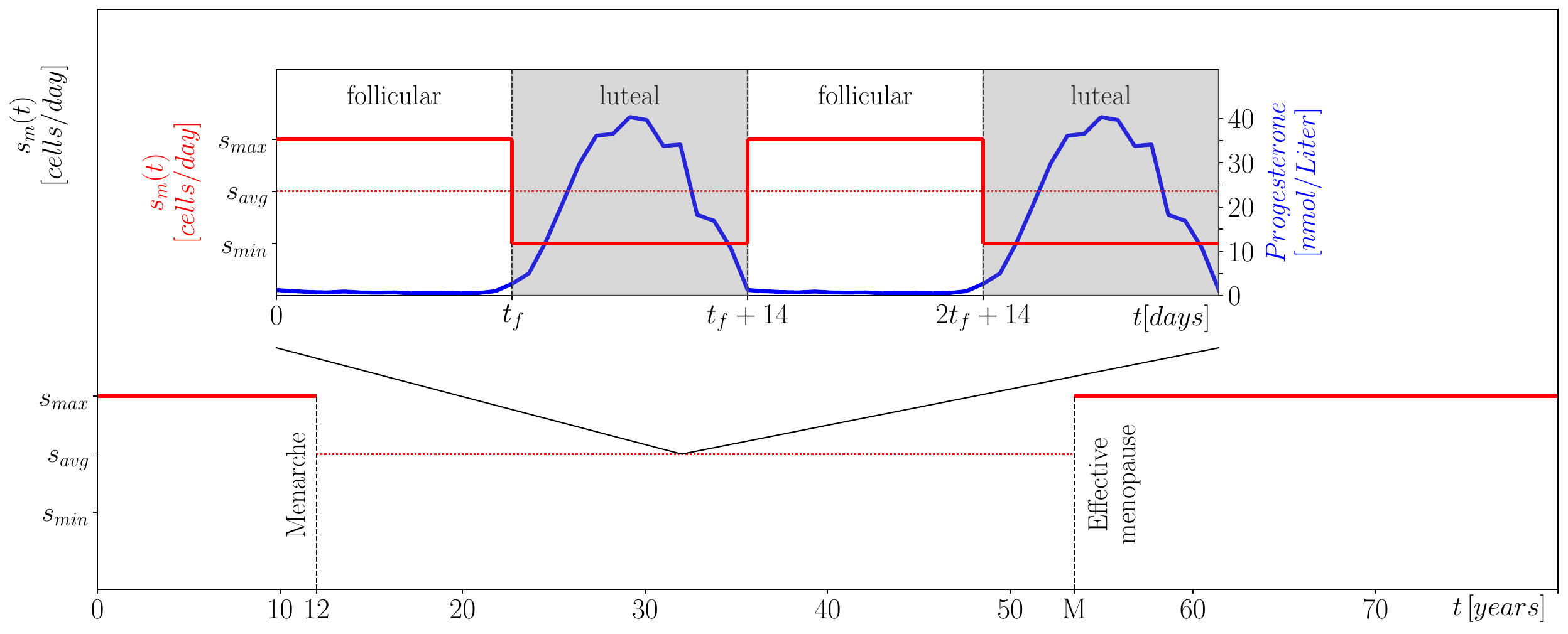}
    \caption{
    The immune response $s(t)$ in Eq.~\eqref{eq:moving_A_def_micro} plotted over the lifetime of a woman. The inset shows (red) the assumed changes in immune response along with (blue) the changes in progesterone levels during the menstrual cycle that occur between the age of 12 (the average age of menarche) and the age of effective menopause $M$. 
    $s_m(t)$ is reduced during the (grey) luteal phase of the  cycle due to the higher levels of progesterone~\cite{mean_prog_cycle,Progesterone_dampens_killer_cells,Progesterone_dampens_immune_responses}.
   $M$ is chosen by the combination of natural menopause and delayed menopause due to hormone replacement therapy; see the three cases in Sec.~\ref{sec:breast_cancer_data}~\ref{sec:combined_model_cases}.
    }
    \label{fig:movingA_global}
\end{figure*}
Alternatively, we can get very good agreement between  $\hat{\mathcal{R}}(t)$ and $\mathcal{R}(t)$ below the age of 50-54,  where we observe a  polynomial increase, at the expense of strong disagreement above the age of 50-54, where there is a change and the increase becomes linear~\cite[Fig. 4]{sup}.
This raises the question: what important process(es) related to breast cancer in women are missing from the model? We address this question in the next section.

\subsubsection{Unusual  age-specific cumulative risk of breast cancer in women: Time-varying immune response}
\label{sec:combined_model_cases}

We now allow the immune response to vary over time to explain the unusual  age-specific cumulative risk of breast cancer in women. 
This results in a moving extinction threshold due to  a combination of a time-varying immune response, $s_m(t)$, and random mutations, ${m}(t)$.

The change in the increase of $\mathcal{R}(t)$ from 
 polynomial to linear occurs around  the age of 44-50~\cite[Fig. 4]{sup}. 
It coincides with the menopause, which begins between the ages of 45 and 55~\cite{51_meno}, when the natural menstrual cycle stops.
This coincidence 
is the first indication that the menstrual cycle plays an important role in the development of breast cancer in women. 
Secondly, there is evidence that higher levels of {\em progesterone} during the luteal phase of the menstrual cycle  weaken the immune system~\cite{abstract_progesterone, Progesterone_dampens_immune_responses, Progesterone_dampens_killer_cells} 
and stimulate the growth of  cancer cells~\cite{Trabert_2020_320}. 
While immunosuppression may be general~\cite{Zwahlen_2024_37} or specific to breast cancer~\cite{Walter_2020_1758547}, biological studies suggest that higher levels of progesterone during the luteal phase have a particularly strong effect on the mammary gland and therefore breast cancer, but not necessarily other cancers in women~\cite{Atashgaran2016}.
It has even been suggested that progesterone is the main driver of breast cancer risk during the menstrual cycle~\cite{Pike1993}.
Thirdly, there is evidence that a shorter menstrual cycle increases the risk of breast cancer~\cite{menstrual_cycle_data_evidence,correspondence_menstrual_cycle_data_evidence,menstrual_cycle_review}. 
These indications lead us to include in the model the effects of varying progesterone levels during the menstrual cycle and, for some women, during menopausal treatment.

The menstrual cycle is made up of different phases:  the {\em luteal phase} lasts on average $14$ days with little variation~\cite{14_day_length_const}, {\em menstruation} lasts on average 5 days, the {\em follicular phase} also lasts on average $14$ days but is much more variable~\cite{var_cycle}, and {\em ovulation} lasts less than 24 hours. The total cycle length is reported to be between 19 and 44 days~\cite{cycle_length}. 

For the model, we assume that menstruation begins at the age of 12, and simplify the menstrual cycle into two phases: a fixed  luteal phase of 14 days~\cite{14_day_length_const}, and a variable follicular phase whose length $t_f$ is normally distributed with a mean of 14 days and a standard deviation of 2.4 days~\cite{var_cycle}. 
We then use the fact that there is a noticeable increase in progesterone levels during the luteal phase~\cite{mean_prog_cycle}, illustrated by the blue curve in the inset of Fig.~\ref{fig:movingA_global}.
There is evidence that this increase weakens the immune response, $s_m$, and may increase the per capita growth rate of cancer cells, $\nu r$.  We propose that the dominant effect 
is a weaker immune response $s_m$ during the luteal phase.\footnote{The approximation~\eqref{eq:approx_stoch_A} shows that an increased per capita growth rate $\nu r$, due to higher  progesterone levels during the luteal phase, has a similar effect to a weaker immune response $s_m(t)$ - both lower the extinction threshold during the luteal phase.}
Specifically, we assume that the immune response in model~\eqref{eq:stochastic_mapping_moving_A} changes as follows during a woman's lifetime:
\begin{equation}
    s_{m}(t)=
    \begin{cases}
    s_{max} & \text{for}\;\; 0 < t < 12,\\
    s_{min} &\text{in luteal phases while}\;\; 12 \le t < M,\\
    s_{max} &\text{in follicular phases while}\;\; 12 \le t < M,\\
    s_{max} & \text{for}\;\; t > M,\\
    \end{cases}
    \label{eq:moving_A_def_micro}
\end{equation}
where $M$ denotes the age of `effective' menopause. This $s_{m}(t)$ is shown in red in the inset of Fig.~\ref{fig:movingA_global}.

Next, we compare the actual age-specific cumulative risk of breast cancer in women obtained from real data, $\mathcal{R}(t)$, with $\hat{\mathcal{R}}(t)$ obtained from  model~\eqref{eq:stochastic_mapping_moving_A} for three different cases of time-varying immune response~\eqref{eq:moving_A_def_micro}.
The model parameters are listed in~\cite[Table 3]{sup}, where some of them are chosen using sparse data on the progression of untreated breast cancer in women~\cite[Sec. 3]{sup}.

{\em Case 1: Average menopause}.
In Fig.~\ref{fig:breast_risk}(b) we use the average age at menopause and set $M=51$ years for all women; see the inset. The (red) $\hat{\mathcal{R}}(t)$ closely reproduces the 
 polynomial increase of the (black) ${\mathcal{R}}(t)$ below the age of 50-54, and shows a distinct change around 50-54 due to menopause. 
However, despite an improvement over the fixed immune response model, the (red) $\hat{\mathcal{R}}(t)$  underestimates  the (black) ${\mathcal{R}}(t)$ above the age of 50-54.

{\em Case 2: Distributed menopause}.
In Fig.~\ref{fig:breast_risk}(c) we use the age distribution at the onset of menopause, that is $M$ is normally distributed with a mean of 51 years and a standard deviation of 4.86 years~\cite{51_meno,4_86_meno_sd}; see the inset. 
The (red) $\hat{\mathcal{R}}(t)$ closely reproduces the polynomial increase of the (black) ${\mathcal{R}}(t)$ below the age of 50-54, shows a less distinct change around 50-54 due to menopause, but still underestimates  the (black) ${\mathcal{R}}(t)$ above the age of 50-54. Although the model is getting closer to the actual data, it is still missing some of the necessary ingredients to achieve a very good agreement.

{\em Case 3: Distributed menopause with hormone replacement therapy (HRT)}.
It turns out that the missing ingredient is HRT. After the menopause, the body naturally stops producing hormones, including progesterone and oestrogen. HRT reduces menopausal symptoms by replacing the natural hormones with synthetic ones so the progesterone changes continue for the duration of HRT. 
Based on the data in~\cite{hrt_Gamma_shape,6_0_mean_HRT}, our model assumes that 26$\%$ of women use HRT, and that their progesterone changes continue past natural menopause for a period of time that has a gamma distribution with a mode of 6 years and a standard deviation of 4.8 years~\cite[Sec. 2.2.5]{sup}.
In Fig.~\ref{fig:breast_risk}(d) we take HRT into account. Specifically, $M$ is first drawn from the normal distribution of the onset of menopause and then, in 26$\%$ of cases, extended by the duration of HRT, which is drawn from the gamma distribution of HRT; see the inset.
The results of the simulations show that the (red) $\hat{\mathcal{R}}(t)$ now closely reproduces the (black) ${\mathcal{R}}(t)$ at all ages.

\subsubsection{Effects of menstrual cycle length on breast cancer risk: Time-varying immune response}

In addition to being able to reproduce the actual age-specific cumulative risk of different cancers,
our complete cancer development model~\eqref{eq:stochastic_mapping_moving_A} with time-varying immune response~\eqref{eq:moving_A_def_micro} naturally accounts for the observation
in~\cite{nearly_double_risk_short_cycle} that the cumulative risk of breast cancer for women with cycles shorter than 25 days is 1.86 times higher than for women with cycles between 25 and 31 days.

\begin{figure}[H]
    \centering
    \includegraphics[width=0.5\textwidth]{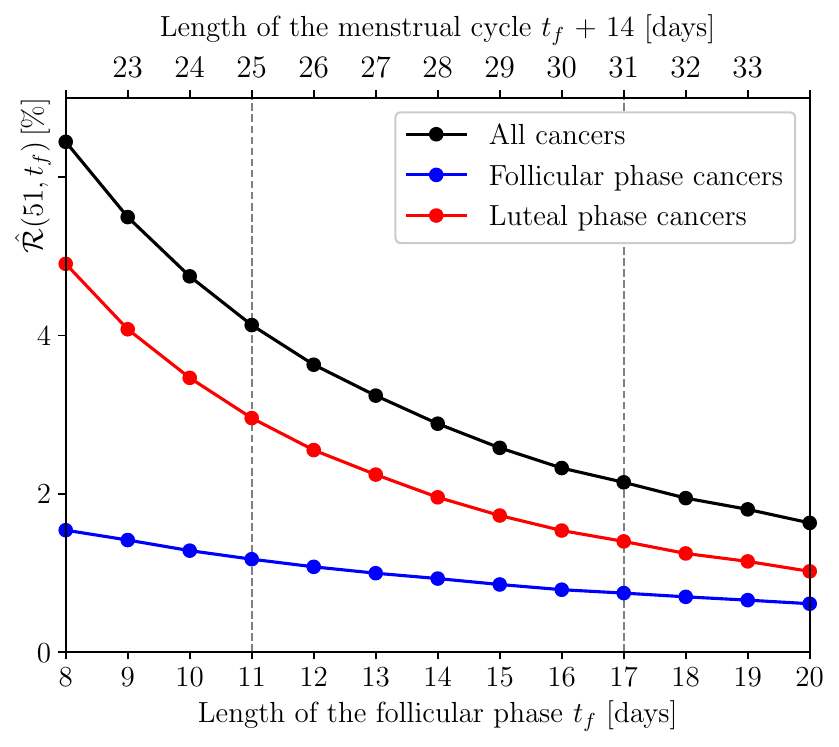}
    \caption{
    (Black) The model's cumulative risk of a woman developing breast cancer by the age of 51, for {\em Case 3} of Sec.~\ref{sec:breast_cancer_data}~\ref{sec:combined_model_cases},  depending on follicular phase length $t_f$ and menstrual cycle length. The length of luteal phase length is fixed at 14 days. (Red and blue) The different risks of  developing breast cancer during the luteal and follicular phases of the cycle, respectively. We use the parameter values in~\cite[Tables 2 and 3]{sup}.
    }
    \label{fig:risk_cycle_length}.
\end{figure}

To show this, we consider {\em Case 3} of Sec.~\ref{sec:breast_cancer_data}~\ref{sec:combined_model_cases}, which gives very good agreement with the real data.
We then examine in Fig.~\ref{fig:risk_cycle_length} how the model's simulated  risk of developing breast cancer by the age of 51, that is $\hat{\mathcal{R}}(51,t_f)$, changes with the length $t_f$ of the much more variable follicular phase, while keeping the luteal phase at 14 days.
Not only does the simulated risk $\hat{\mathcal{R}}(51,t_f)$ decrease significantly with increasing follicular phase length and thus menstrual cycle length, but it also predicts a 1.75-fold increase in risk for women with cycles between 22 and 25 days compared to women with cycles between 25 and 31 days. This prediction is in very good agreement with the observation in~\cite{nearly_double_risk_short_cycle}.

The model also shows that the greater proportion of transitions to cancer occur during the luteal phase when the progesterone levels are higher and the immune response is weaker. This proportion ranges from 76\% for an 8-day follicular phase to 63\% for a 20-day follicular phase, as shown by the red and blue curves, respectively, in Fig.~\ref{fig:risk_cycle_length}.
Thus, it may very well be that progesterone is the main driver of breast cancer risk during the menstrual cycle, as suggested in~\cite{Pike1993}.
We refer to~\cite[Sec.2.2.6]{sup} for more details.

\section{Conclusions and Outlook}
\label{sec:conclusion}

In the context of cancer development, we define an extinction threshold as a critical size of a cluster of mutated cells below which the cluster can be eliminated by the immune system alone, and above which the cluster develops into cancer.
We then construct a simple dynamic model to demonstrate that an extinction threshold is an essential component of cancer development as evidenced by very good agreement with a wide range of real cancer data. 
The model incorporates random mutations, deterministic growth of mutated cells, and an immune response that eliminates mutated cells. An important feature of the model is that the extinction threshold can move over time to reflect changes in the immune response, or other cancer-related processes, due to seasonal variations, viral infections,  the menstrual cycle, immune deficiencies, immunotherapy, ageing, etc.

We started with the defining feature of cancer progression: an asymmetry in which small cancers grow faster than large cancers saturate as they approach carrying capacity. This allowed us to identify the limitations of classical extinction threshold models from population biology in reproducing cancer progression, and to derive a new model that overcomes these limitations.
We then showed that the new model accurately reproduces a wide range of real-world cancer data, from the progression of breast cancer in mice to the population-level age-specific cumulative risk of different cancers in humans.

While we believe  that this model can be applicable to a wide variety of cancers, we
have chosen to focus on breast cancer for a number of reasons. Firstly, breast cancer is 
 one of the most common cancers in the world.
Secondly, breast cancer in women has an unusual age-specific cumulative risk that appears to 
increase  polynomially up to the age of menopause and linearly thereafter,  which differs from the exponential increase across all ages for many other cancers.
Thirdly, this difference has not been fully understood in terms of the underlying dynamics of cancer development. 
To the best of our knowledge, this is the first report on the prediction of actual cancer development and age-specific cumulative risk using dynamic models with an extinction threshold.
We have shown that the concept of a moving extinction threshold, which reflects the different immune response at different phases of the menstrual cycle and hormone replacement therapy (HRT), can explain the unusual age-specific cumulative risk of breast cancer in women. 
In addition, our results
are consistent with the observations that the risk of breast cancer is higher in women with shorter menstrual cycles and that combined 
HRT  with oestrogen and progesterone increases the cumulative risk of breast cancer~\cite{Breast_cancer_HRT,Breast_cancer_HRT_and_Combined,whi_2002}.
The latter observation led to a debate about the benefits of combined HRT versus the increased risk of cancer. 

Our model could make a valuable contribution to this debate by providing new insights into the effects of combined HRT;  note the difference between (red) $\hat{\mathcal{R}}(t)$ in panels (c) and (d) in Fig.~\ref{fig:breast_risk}.  In particular, it provides a basis for development of more detailed models to better quantify the effects of HRT on breast cancer risk in women.
More generally, the moving extinction threshold approach can be applied to a variety of other cancer scenarios where the immune response or other threshold parameters, such as the per capita growth rate of cancer cells, vary over time. 

From the point of view of mathematical modelling,
recently there has been much interest in the theory of tipping points (critical transitions)~\cite{ashwin2012} and evolutionary games for adaptive therapy of cancer~\cite{adaptive_therapy}. 
It turns out that carcinogenesis in our model is an example of a noise-induced tipping point: a point in time when random mutations cause a critical transition across the extinction threshold from a cancer-free state  to the first stage of cancer.
At the same time, our model of cancer development is simple enough for the techniques of evolutionary games and optimal adaptive therapy~\cite{Gluzman_opt_cancer,opti_control_stoch_vladimirsky}.  These features make our model a natural candidate for combined analysis using the techniques of tipping point theory and evolutionary games to explore new optimal cancer treatment strategies.
Furthermore, the model can be easily extended to capture more processes within the tumour microenvironment by adding separate equations governing the dynamics of the immune cells~\cite{kuznetsov1994nonlinear} and the tumour stroma~\cite{noemi_logistic}.

\bibliographystyle{unsrt}
\bibliography{references.bib}

\end{document}


\maketitle

\section{Technical details for Section 2 
}

\subsection{Classical growth models with no extinction
threshold}
\label{sec:richard_appendix}
In this section, we linearise Richards' model and show  how the defining feature of cancer  progression depends on the shape parameter $\nu$. Richards' model 
\begin{equation}
    \frac{dx}{dt} = \nu r x \left[1-\left(\frac{x}{K}\right)^\frac{1}{\nu}   \right]=f(x),
    \label{eq:generalised_logistic_app}
\end{equation}
gives the logistic equation for $\nu=1$ and Gompertz' equation in the limit $\nu\to\infty$,
\begin{equation*}
 \lim_{\nu\to\infty} f(x) =
    \lim_{\nu\to\infty} \nu \left[1-\left(\frac{x}{K}\right)^\frac{1}{\nu}\right] = -\ln\left(\frac{x}{K} \right).
\end{equation*}
Next, we linearise Eq.\eqref{eq:generalised_logistic_app}:
\begin{equation*}
   \frac{df}{dx}(x) =r\nu-r(\nu+1) \left(\frac{x}{K}\right)^{\frac{1}{\nu}}.
\end{equation*}
Inserting the equilibria $x = 0$ and $x =  K$ into the linearisation above gives the following eigenvalues
\begin{equation*}
    \begin{split}
        \lambda_{grow} &= \left.\frac{df}{dx}\right\vert_{x=0} = r\nu,\\
        \lambda_{sat} &= \left.\frac{df}{dx}\right\vert_{x=K} = -r,
    \end{split}
\end{equation*}
and the defining feature 
$$
\left|\frac{\lambda_{grow}}{\lambda_{sat}}\right|=\nu.
$$
\subsection{Model 1: A classical growth model with an extinction threshold}
\label{sec:gen_proposition}
In this section, we consider Model 1 given by
\begin{equation}
    \begin{split}
        \dot{x}
        &=-\nu^2 r x  + \nu^2 \mu x^\frac{\nu + 1}{\nu}- \nu^2 \gamma x^\frac{\nu + 2}{\nu}\\
        &=-\nu^2 r x 
        \left(1-\left(\frac{x}{A}\right)^{\frac{1}{\nu}}\right)
        \left(1-\left(\frac{x}{K}\right)^{\frac{1}{\nu}}\right)\\
        &=f(x),
    \end{split}
    \label{eq:gen_Allee_app}
\end{equation}
where we choose
\begin{equation}
\label{eq:M1par}
0 < \gamma < \mu^2/(4r),
\end{equation}
to ensure that the system has two positive equilibra
\begin{equation}
\label{eq:AK}
A=\left(\frac{\mu - \sqrt{\mu^2 - 4 r\gamma}}{2\gamma}\right)^{\nu}, 
\quad
K =\left(\frac{\mu + \sqrt{\mu^2 - 4 r\gamma}}{2\gamma}\right)^{\nu}
\quad\mbox{such that}\quad 
0 < A < K.
\end{equation}
Condition~\eqref{eq:M1par} is derived by writing the equilibrium equation in terms of $y=x^{1/\nu}$ as
$$
0 = -\nu^2 x 
\left(
\gamma y^2 - \mu y + r
\right)
$$
and requiring that the quadratic function of $y$ in the bracket has two distinct real roots.

We first consider two special cases of $\nu$ and then
prove that, if $\nu > 0$, small populations grow slower than large populations saturate as they approach carrying capacity $K$.

When $\nu=1$, Model 1 reduces to the Volterra's equation~\cite{Volterra} with
$$
f(x) = -r x  + \mu x^2 -  \gamma x^3 = 
-r x \left(1-\frac{x}{A}\right)
        \left(1-\frac{x}{K}\right).
$$
In the limit $\nu \to \infty$ 
Model 1 converges to the Gompertz model with an Allee effect proposed by Amarti~{\em et al} ~\cite{gompertz_allee}. To see that, note that
\begin{equation}
    \begin{split}
        &\lim_{\nu\to\infty}-\nu^2 r x 
        \left(1-\left(\frac{x}{A}\right)^{\frac{1}{\nu}}\right)
        \left(1-\left(\frac{x}{K}\right)^{\frac{1}{\nu}}\right)\\
        =&\lim_{\nu\to\infty}
        \nu r x
        \left(1-\left(\frac{x}{K}\right)^{\frac{1}{\nu}}\right)
        \lim_{\nu\to\infty}-\nu
        \left(1-\left(\frac{x}{A}\right)^{\frac{1}{\nu}}\right),
    \end{split}
    \label{eq:limit_volterra}
\end{equation}
where
\begin{equation*}
    \begin{split}
        &\lim_{\nu\to\infty}\nu r x
        \left(1-\left(\frac{x}{K}\right)^{\frac{1}{\nu}}\right)
        =-r x\ln\left(\frac{x}{K}\right),\\
        &\lim_{\nu\to\infty}-\nu
        \left(1-\left(\frac{x}{A}\right)^{\frac{1}{\nu}}\right)
        =\ln\left(\frac{x}{A}\right),
    \end{split}
\end{equation*}
which gives
\begin{equation}
    \lim_{\nu\to\infty}f(x)=
    -r x\ln\left(\frac{x}{K}\right)\ln\left(\frac{x}{A}\right).
    \label{eq:volterra_limit}
\end{equation}

Next we examine the defining feature of cancer progression in Model 1 with parameters satisfying~\eqref{eq:M1par}. The equilibria of Eq.\eqref{eq:gen_Allee_app} are given by
\begin{equation*}
        x=0,\quad
        x=A \quad\mbox{and}\quad
        x=K,
    \label{eq:eq_gen_Allee_app}
\end{equation*}
and the linearisation is given by
\begin{equation}
    \begin{split}
        \frac{df}{dx} (x) =&
        \nu  r
        \bigg(\left(-(\nu +2) \left(\frac{x}{A}\right)^{1/\nu }+\nu +1\right) \left(\frac{x}{K}\right)^{1/\nu }\\
        &+\nu  \left(\frac{x}{A}\right)^{1/\nu }+\left(\frac{x}{A}\right)^{1/\nu }-\nu \bigg).
    \end{split}
\end{equation}
Inserting the equilibria into the linearisation above yields:
\begin{equation}
    \begin{split}
        \lambda_0 &=
        \left.\frac{df}{dx}\right\vert_{x=0} =
        - r\nu^2,\\
        \lambda_{grow}
        &=
        \left.\frac{df}{dx}\right\vert_{x=A}
        =
        r\nu\,
        \left(1-\left(\frac{A}{K}\right)^{\frac{1}{\nu}}\right),\\
        \lambda_{sat}
        &=
        \left.\frac{df}{dx}\right\vert_{x=K}
        =
         r\nu\,
        \left(1-\left(\frac{K}{A}\right)^{\frac{1}{\nu}}\right).
    \end{split}
    \label{eq:eq_gen_Allee_eigenvalues}
\end{equation}
We now have the necessary ingredients to prove that small populations always grow slower than large populations saturate at $K$.
\begin{proposition}
\label{sec:proposition_gen_volt}
    Consider Model 1 given by Eq.~\eqref{eq:gen_Allee_app} in the parameter region~\eqref{eq:M1par} where there are two positive equilibria $A$ and $K$  given by Eq.~\eqref{eq:AK}. 
    Then, the defining feature satisfies
    \begin{equation*}
    0 < \left|\frac{\lambda_{grow}}{\lambda_{sat}}\right| < 1
     \quad\mbox{and}\quad
      \lim_{K\to\infty} \left|\frac{\lambda_{grow}}{\lambda_{sat}}\right| = 0
    \quad\mbox{for}\quad \nu> 0.
    \end{equation*}
\end{proposition}
\begin{proof}
From~\eqref{eq:eq_gen_Allee_eigenvalues}, we have
    \begin{equation}
    \label{eq:def_feat}
        \frac{\lambda_{grow}}{\lambda_{sat}} = 
        \frac{\left(1-(A/K)^{1/\nu}\right)}{\left(1-(K/A)^{1/\nu}\right)} =
        \left(\frac{A}{K}\right)^{1/\nu} \frac{\left(1 -(A/K)^{1/\nu}\right)}{\left((A/K)^{1/\nu} - 1\right)} = -  \left(\frac{A}{K}\right)^{1/\nu}. 
    \end{equation}
Next, we note that since $0 < A < K$, the defining feature $|\lambda_{grow}/\lambda_{sat}|$ is a monotone increasing function of $\nu$ for all $\nu > 0$. Furthermore, we have that 
    \begin{equation*}
    \lim_{\nu\to 0} 
    \left(\frac{A}{K}\right)^{1/\nu} 
    = 0 
    \quad \mbox{and} \quad
    \lim_{\nu\to\infty}
     \left(\frac{A}{K}\right)^{1/\nu}   
        = 1.
    \end{equation*}
Thus, $|\lambda_{grow}/\lambda_{sat}|$ is bounded between 0 and 1.
The large $K$ limit follows directly from~\eqref{eq:def_feat} for any fixed $\nu > 0$.
\end{proof}

\subsection{Model 2: A new growth model with an extinction threshold}
\label{sec:deterministic_A}

In this section, we consider Model 2 with a constant $s(t)=s$, given by
\begin{equation}
    \frac{dx}{dt} =
    \begin{cases}
        \nu r x - \nu \mu x^{\frac{\nu+1}{\nu}} - s = f(x)& \text{ if }x>0,\\
        0&\text{ if }x=0,
    \end{cases}
    \label{eq:richards_harvest_appendix}
\end{equation}
where we choose
\begin{equation}
\label{eq:M2par}
    0 < s <
 \frac{1}{\mu^\nu} 
 \left(
\frac{\nu r}{\nu + 1}
 \right)^{\nu + 1},
\end{equation}
to ensure that the model has two positive equilibria, $A$ and $K$, such that $0 < A < K$.
The upper bound in condition~\eqref{eq:M2par} corresponds to the value of $s$ at which equilibria $A$ and $K$ meet in a saddle-node (fold) bifurcation and subsequently disappear, that is, when $A = K$. It was derived by setting 
\begin{equation}
    \begin{split}
        f(x) & =  \nu r x - \nu \mu x^{\frac{\nu+1}{\nu}} - s = 0,\\
        f'(x)& =  \nu  r - \mu (\nu + 1)  x^{\frac{1}{\nu }}= 0,
    \end{split}
\end{equation}
then eliminating $x$ to obtain
\begin{equation}
    s = \frac{1}{\mu^\nu}\left(\frac{\nu r}{\nu + 1} \right)^{\nu + 1}.
\end{equation}

We will now approximate the location of the extinction threshold $A$.
We introduce the rescaled time and parameters.
\begin{equation*}
\tau=\frac{t}{\nu r},\quad
        \Tilde{\mu}= \frac{\mu}{r},\quad\text{and}\quad
        \Tilde{s} = \frac{s}{\nu r},
\end{equation*}
and rewrite the first component of Model 2 as follows
\begin{equation}
    \frac{dx}{d\tau} =x - \Tilde{\mu} x^{\frac{\nu +1}{\nu}} - \Tilde{s}.
    \label{eq:rescaled_richards_harvest}
\end{equation}
We 
then write the equilibrium condition,
\begin{equation*}
    0 = x - \Tilde{\mu} x^{\frac{\nu +1}{\nu}} - \Tilde{s},
\end{equation*}
and consider two cases.

{\bf Case I}: When $\nu=1$, we have
\begin{equation*}
    0 = x - \Tilde{\mu} x^{2} - \Tilde{s},
\end{equation*}
two positive equilibria
\begin{equation*}
    A=\frac{1-\sqrt{1-4\Tilde{\mu}\Tilde{s}}}{2\Tilde{\mu}}, 
    \quad\mbox{and}\quad
    K=\frac{1+\sqrt{1-4\Tilde{\mu}\Tilde{s}}}{2\Tilde{\mu}},
\end{equation*}
and are interested in the extinction threshold $A$. If we additionally assume that
$$
4\Tilde{\mu}\Tilde{s} \ll 1
\quad\mbox{or}\quad
4\mu s \ll  r^2,
$$
we can set 
$$
\epsilon = 4\Tilde{\mu}\Tilde{s},
$$
expand the formula for $A$ in different powers of $\epsilon$,
\begin{equation}
    A = 
    \frac{1-\sqrt{1 - \epsilon}}{2\Tilde{\mu}}
    =
    \frac{\epsilon}{4 \Tilde{\mu}} + \frac{\epsilon^2}{16\Tilde{\mu}}+\mathcal{O}(\epsilon^3),
\end{equation}
and approximate the extinction threshold with an error $\mathcal{O}(\epsilon^2)$ as
\begin{equation*}
    A = \frac{\epsilon}{4 \Tilde{\mu}} = \Tilde{s} =\frac{s}{\nu r}.
\end{equation*}

{\bf Case II}: When $\nu\gg1$, we have
\begin{equation}
    0=x - \Tilde{\mu} x^{\frac{\nu +1}{\nu}} - \Tilde{s}.
\end{equation}
We set 
$$
\epsilon=1/\nu\ll1,
$$ 
and  write the nonlinear term as follows
\begin{equation*}
    x^{\frac{\nu+1}{\nu}}=x x^\frac{1}{\nu}=xx^\epsilon,
\end{equation*}
where
$$
x^\epsilon = e^{\ln (x^{\epsilon})} = e^{\epsilon\ln (x)},
$$
can be expanded in different powers of $\epsilon$,
\begin{equation*}
    x^\epsilon = 1 + \ln (x)\, \epsilon + \frac{1}{2}\ln (x)^2\, \epsilon^2 + \mathcal{O}(\epsilon^3),  
\end{equation*}
to obtain
\begin{equation}
\label{eq:ee}
    0 =x - \Tilde{\mu} x\left(1+\frac{\ln x}{\nu} + \frac{\ln(x)^2}{2\nu^2}+\mathcal{O}(\epsilon^3)\right) - \Tilde{s}.
\end{equation}
If we additionally assume that 
$$
\nu^2 \gg \ln(A)^2 > 1,
$$
we can approximate Eq.~\eqref{eq:ee} near $x=A$ with an error $\mathcal{O}(\epsilon^2)$ as
\begin{equation}
\label{eq:xA}
    0 =  x - \Tilde{\mu} x\left(1+\frac{\ln x}{\nu}\right) - \Tilde{s}.
\end{equation}
We then set $x=A$ and additionally substitute $A= 10^a$ for some
$a > 0$  in the argument of the logarithm to obtain

\begin{equation*}
        0 =
        A -\Tilde{\mu}\left(1+\frac{a\ln 10}{\nu}\right) A -\Tilde{s},
\end{equation*}
yielding the following approximation of the extinction threshold
\begin{equation*}
    A= \frac{s}{\nu r-\mu(a\ln 10+\nu)}.
\end{equation*}
Therefore, the extinction threshold of Eq.~\eqref{eq:richards_harvest_appendix} can be approximated as
\begin{equation*}
    A\approx
    \begin{cases}
        s/(\nu r)&\text{if}\quad \nu = 1 \;\;\text{and}\;\; 4 \mu s \ll r^2, \\
        s/(\nu r - \mu (a\ln 10+\nu))&\text{if}\quad \nu^2\gg\ln (A)^2 > 1\;\; \text{and}\;\; A\sim10^a.
    \end{cases}
\end{equation*}
\subsection{Carcinogenesis in the presence of an extinction threshold}

\begin{figure}[th]
    \centering
    \includegraphics[width=\textwidth]{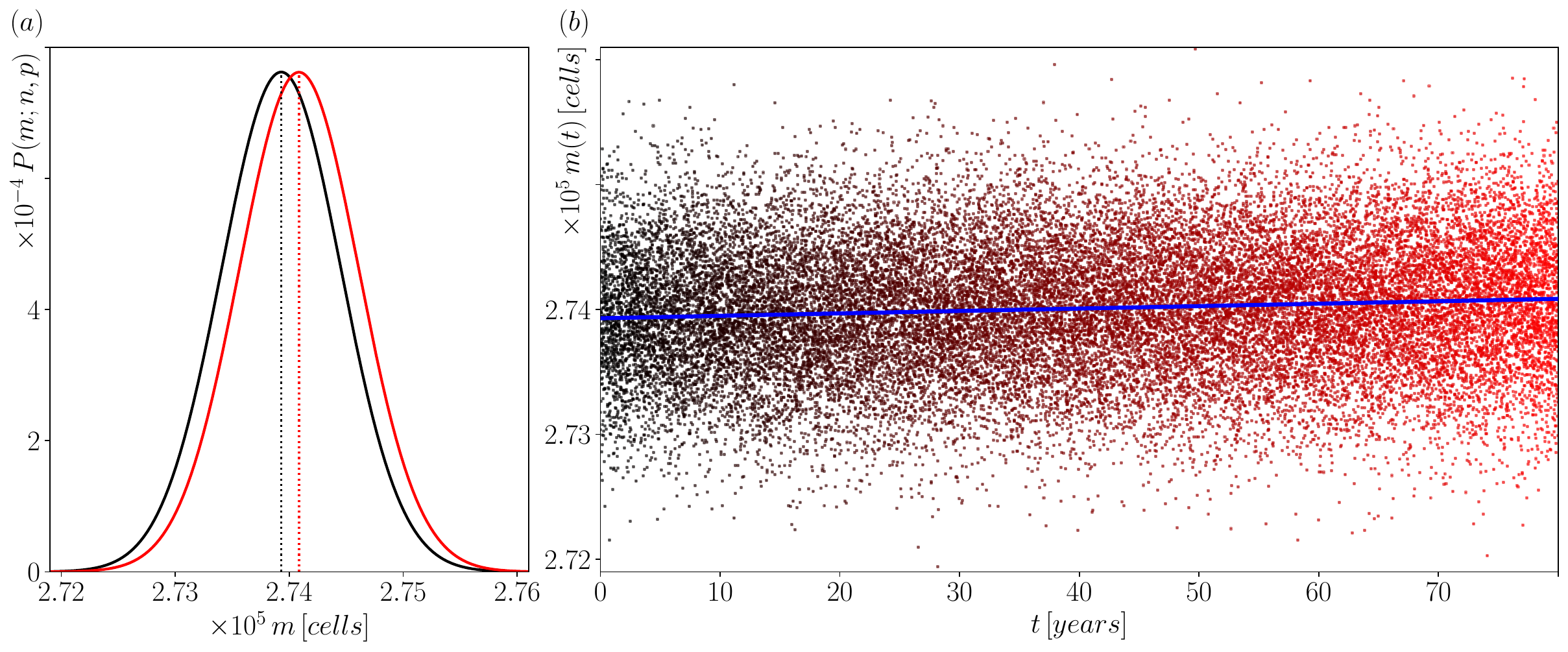}
    \caption{
    (a) The binomial probability distribution~\cite[Eq. (2.8)]{main_paper} for $m$  new mutations from $n=10^{10}$ stem cells in one day is shown at two points in the lifetime: (black) at birth with $p(0) =p_0 \approx \SI{2.739e-5}{}$ and (red) at life expectancy $T=80$ years with  $p(T)\approx\SI{2.741e-5}{}$.
    (b) (Black-red dots) A corresponding example of a single realisation of new mutations per day, $m(t)$, drawn from the binomial distribution~\cite[Eq. (2.8)]{main_paper} with $p(t)$ slowly increasing from $p_0$ to $p(T)$ according to~\cite[Eq. (2.9)]{main_paper}) with $\delta=0.06\%$. 
    ((Blue line) The slowly varying expected value of new mutations in cells per day $\langle m \rangle(t) = np(t)$.
    }
    \label{fig:2_binomialDistributions}
\end{figure}

Figure~\ref{fig:2_binomialDistributions} shows the effect of the slowly increasing probability $p(t)$ in the binomial distribution in~\cite[Eq. (2.8)]{main_paper}. 

\subsection{The complete cancer development model}
\subsubsection{Extinction threshold in the complete cancer development model}
\label{sec:noisy_A}

In this section, we approximate instantaneous position of the {\em moving extinction threshold} $\bar{A}(t)$ in the complete development model~\cite[Eq.~(2.10)]{main_paper} with
\begin{equation*}
    F(x,t) = x(t) + \nu r x(t) - \nu \mu x(t)^{\frac{\nu+1}{\nu}} - s_m(t) + m(t).
\end{equation*}
First, we take expected values and consider the  equation for the fixed point of $\langle  F(x,t) \rangle$,
$$
\langle F(x,t) \rangle - \langle x \rangle = 0.
$$
We then use Eq.~\eqref{eq:xA} to write an approximation to the fixed point equation near $x=A$, 
\begin{equation*}
 \nu r \langle x \rangle -\nu \mu \langle x \rangle \left(1+\frac{\langle \ln x \rangle}{\nu}\right) - \langle s_m(t) \rangle + \langle m(t) \rangle
 = 0.
\end{equation*}
We then set $x=A$ and additionally substitute $A = 10^a$ for some
$a > 0$  in the argument of the logarithm to obtain
$$
     \nu r \langle {A}  \rangle- \nu \mu\langle {A} \rangle \left(1+\frac{a\ln 10}{\nu}\right)-\langle s_m(t)\rangle + \langle m(t)\rangle
     = 0.
$$  
Since $s_m(t)$ is constant within each phase of the menstrual cycle (see \cite[Eq.~(3.1)]{main_paper}), we can restrict to individual phases and write
$$
\langle s_m(t)\rangle = s_m(t).
$$
Furthermore, we replace the expected values of $A$ and $m(t)$ with their time averages in each phase of the menstrual cycle
$$
\bar{A}(t) = \langle A \rangle 
\quad\mbox{and}\quad
\overline{m}(t) = \langle m(t)\rangle.
$$
This gives us
$$
\nu r  \bar{A}(t)  - \nu \mu \bar{A}(t)  \left(1+\frac{a\ln 10}{\nu}\right) -  s_m(t) + \overline{m}(t)
= 0,
$$
which leads to~\cite[Eq. (2.11)]{main_paper},
\begin{equation}
\label{eq:avA}
\bar{A}(t) = \frac{s_m(t) - \overline{m}(t)}{\nu r-\mu(\nu+a\ln 10)}.
\end{equation}

\subsubsection{Rescue events in the complete cancer development model}
\begin{figure}[th]
    \centering
         \includegraphics[width=0.7\textwidth]{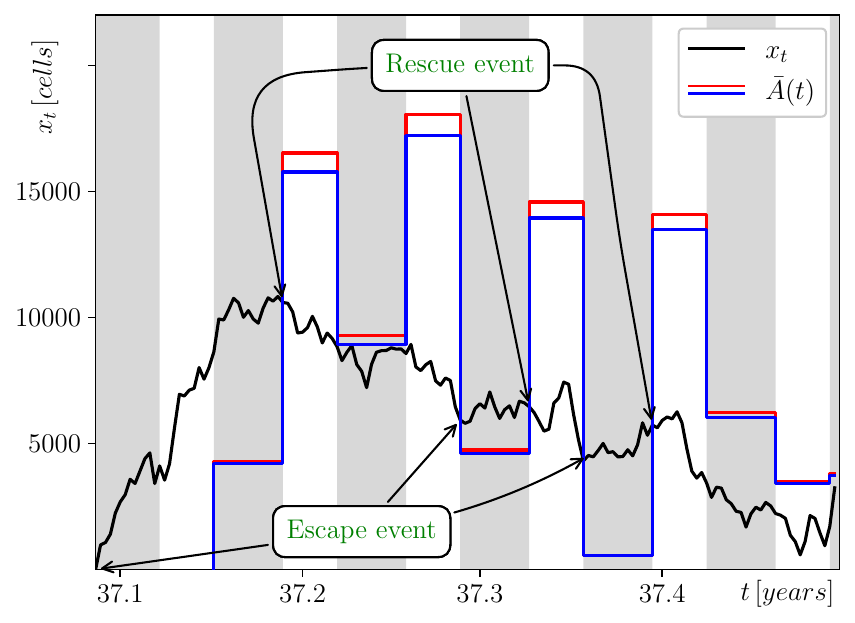}
    \caption{
    An example of breast cancer development $x(t)$ in model~\cite[Eq. (2.10)]{main_paper} for a single realisation of random mutations $m(t)$ using the same parameter values as \cite[Fig.~4]{main_paper} (see also Tab.~\ref{tab:parameter_table_human})
    and for periodic variations in the immune response $s_m(t)$ due to menstrual cycle in~\cite[Eq. (3.1)]{main_paper} with $s_m(t)$ taking values: (grey stripes) $s_{min}= np_0 + 262$  cells/day 
    during  the 14-day long luteal phases with a weaker immune response, and (white stripes) $s_{max}= np_0 +354$ cells/day during  the 11-day long follicular phases with a stronger immune response.
    The (black) number of mutated cells $x(t)$ exceeds the (blue/red) moving extinction threshold $\bar{A}(t)$ during "escape events" and subsequently falls again  below the the (blue/red) moving extinction threshold during "rescue events".
    In each phase, the time average $\overline{m}$ of $m(t)$ is used to obtain $\bar{A}(t)$ (red) numerically and (blue) using the approximation in Eq.~\eqref{eq:avA}.
    }
    \label{fig:RescueEvents}
\end{figure}

Figure~\ref{fig:RescueEvents} shows examples of rescue events in a solution to the complete cancer development model~\cite[Eq. (2.10)]{main_paper}

\section{Technical details for Section 3}

In Sec.~\ref{sec:mice_fit} we present the details of fitting the cancer progression models to mice data.
Sec.~\ref{sec:cum_risk} provides the details for \cite[Sec. 3(b)]{main_paper}.
In particular, we illustrate how the actual age-specific cumulative risk, denoted $\mathcal{R}(t)$, is estimated from age-specific incidence data in Sect.~\ref{sec:cum_risk_data}, and how the theoretical age-specific cumulative risk, denoted $\hat{\mathcal{R}}(t)$, is obtained from the complete cancer development model in Sec.~\ref{sec:cum_risk_model}.
Details of stochastic simulations are provided in Sec.~\ref{sec:tech_detail}, and in Sec.~\ref{sec:fixed_immune} parameters for simulations of colorectal cancer are presented.
In Sec.~\ref{sec:HRT}, we explain how we model the duration of hormonal replacement therapy.
Finally, in Sect.~\ref{sec:cycle_risk_par_expl}, we describe how the relationship between the cumulative risk of breast cancer and the length of the menstrual cycle was estimated.

\subsection{Breast cancer progression in mice: A
comparison between Models and data}
\label{sec:mice_fit}

In order to fit Model 1 and 2 to mice progression data, we consider solutions to the initial value problem (IVP)
\begin{equation*}
    \dot{x} = f(x; \beta),\;\; x(t_0) = x_0,
\end{equation*}
where $x_0$ is the initial breast cancer cell count, injected at the start of the experiment, and $\beta$ is the set of the model parameters to be estimated. Note that $\beta=(r, \nu,A, K)$ for Model 1 and $\beta=(r, \nu, \mu, s)$ for Model 2.

The best fit is computed by minimising the residual sum of squares (RSS)~\cite{mlr_bible} between the solution $x(t)$ of the IVP and the actual values:
\begin{equation}
    RSS = \sum_{i=1}^N (x_i-x(t_i))^2,
    \label{eq:rss}
\end{equation}
where $N$ is the number of data points and $x_i$ is the number of breast cancer cells estimated from the reported tumour volume data at time $t_i$ by assuming a spherical tumour size and a cancer cell density of $10^9$ cells/cm$^3$~\cite{1cm3_cell_count}.
Furthermore, we note that for the data of Vaghi {\em et. al.}~\cite{mice_data_cloud} each grey dot represents a measurement for an individual mouse, whereas Cabeza {\em et. al.}~\cite{treatment_mice} reported the average across several mice.
The minimisation problem \eqref{eq:rss} was solved using the fitting library Non-Linear Least-Squares Minimization and Curve-Fitting for Python (LMFIT) which implements the Levenberg–Marquardt algorithm~\cite{lmfit}.

\subsection{Age-specific cumulative risk of cancer
}
\label{sec:cum_risk}

\subsubsection{Age-specific cumulative risk obtained from age-specific incidence data}
\label{sec:cum_risk_data}

In \cite[Sec 3]{main_paper}, we adopt the approach of Armitage and Doll~\cite{ArmitageDoll} and fit our  complete cancer development model to the  actual age-specific cumulative risk of breast cancer.
To estimate the actual age-specific cumulative risk of cancer from age-specific incidence data, we follow~\cite{hazard_rate_book}.

The cumulative risk at a given age $t$, denoted $\mathcal{R}(t)$, is the probability that cancer develops between ages $0$ and $t$.
When viewed as a function of $t$, $\mathcal{R}(t)$ is the age-specific cumulative risk. 
The incidence rate at a given age $t$, denoted $\rho(t)$, is the probability rate (or per capita rate) that cancer develops at age $t$. When viewed as a function of $t$, $\rho(t)$ is the age-specific incidence rate. 
The time integral of $\rho(u)$ over $u\in[t_1,t_2]$ is the ratio of cancer cases to the entire population during the time interval $[t_1,t_2]$.

To express $\mathcal{R}(t)$ in terms of $\rho(t)$ it is useful to introduce  the age $T$ at which cancer may develop, and the probability 
\begin{equation}
    Pr(t_1 < T < t_2)
    \label{eq:prob_t1t2}
\end{equation}
that cancer develops between ages $t_1$ and $t_2$.
Accordingly, the cumulative risk at time $t$,
\begin{equation}
\mathcal{R}(t) =  Pr(0<T < t),
\label{eq:cumRiskAsProb}
\end{equation}
is the probability that cancer develops between ages 0 and $t$.
Additionally, we introduce the following conditional probabilities:  
$$
Pr(t < T \le t + \Delta t \, | \, T > t)
$$
which denotes the probability that cancer develops during the time interval $[t, t + \Delta t]$, given that no cancer develops up to time $t$,  and 
$$
Pr( T > t\, |\, t< T\leq t+\Delta t),
$$
which denotes the probability that cancer develops at some point after time $t$, given that it has developed in the time interval $[t, t+\Delta t]$.
We note that, by definition,  this second conditional probability is always 1, namely
\begin{equation}
    Pr( T > t\, |\, t< T\leq t+\Delta t) = 1.
    \label{eq:conditinalThatItOccured}
\end{equation}
Finally, we introduce the probability that cancer does not develop before age 
$t$,
\begin{equation}
Pr(T > t) = 1- \mathcal{R}(t).
\label{eq:survival_prop}
\end{equation}

Now, we have all the ingredients to define $\rho(t)$ in terms of a conditional probability as follows~\cite{hazard_rate_book},
\begin{equation}
   \rho(t) = \lim_{\Delta t \to 0} \frac{Pr(t<T\leq t+\Delta t\;|\; T > t)}{\Delta t}.
\end{equation}
Next, we use Bayes theorem to write
\begin{equation}
    \rho(t) = \lim_{\Delta t \to 0} \frac{Pr( T > t\;|\; t<T\leq t+\Delta t)\;Pr (t<T\leq t+\Delta t)}{Pr(T> t) \Delta t},
    \label{eq:bayes}
\end{equation}
and Eqs.~\eqref{eq:conditinalThatItOccured} and \eqref{eq:survival_prop} to write
\begin{equation}
    \rho(t) = \frac{1}{1-\mathcal{R}(t)} \lim_{\Delta t \to 0} \frac{Pr (t<T\leq t+\Delta t)}{\Delta t}.
    \label{eq:bayes_intermediate}
\end{equation}
We then note that $T$ is in $[t, t+\Delta t]$ if and only if $T$ is in $[0,t+\Delta t]\setminus [0,t)$, and use the addition rule for probabilities together with 
 \eqref{eq:cumRiskAsProb} to write: 
\begin{equation}
    Pr (t<T\leq t+\Delta t) = Pr(0<T\leq t+\Delta t) - Pr(0<T\leq t)
    = \mathcal{R}(t+\Delta t) - \mathcal{R}(t).
\end{equation}
We substitute this result into \eqref{eq:bayes_intermediate} to obtain: 
\begin{equation}
    \begin{split}
       \rho(t) &= \frac{1}{1-\mathcal{R}(t)} \lim_{\Delta t \to 0} \frac{\mathcal{R}(t+\Delta t)- \mathcal{R} (t)}{\Delta t},
    \end{split}
\end{equation}
and use the definition of a derivative to obtain
\begin{equation}
    \begin{split}
       \rho(t) &=  \frac{1}{1-\mathcal{R}(t)} \frac{d \mathcal{R}(t)}{dt}
        = -\frac{d}{dt}\ln(1-\mathcal{R}(t)) 
        .
    \end{split}
\end{equation}
Integrating both sides with respect to $t$ and rearranging for $\mathcal{R}(t)$, we arrive at the relation between the age-specific cumulative risk and age-specific incidence rate: 
\begin{equation}
    \mathcal{R}(t) = 1 - e^{-\int_0^t \rho (u) du}.
    \label{eq:cum_risk_derivation}
\end{equation}

Cancer registries such as the National Cancer Registry Ireland (NCRI)~\cite{NCRI}, attempt to measure $\rho(t)$.
For practical reasons, the observed population is grouped into age cohorts depending on the age of individuals.
We use $a_i$ to denote a specific age-cohort for age $i$ up to $i+5$, containing all individuals whose age $t\in [i ,i+5)$:
\begin{equation*}
    a_i = [i,  i+5) \text{ for }i=0, 5, \dots, 75.
\end{equation*}
For each age-cohort $a_i$, the NCRI\cite{NCRI} estimates and reports an age-specific incidence rate $\rho(a_i)$.
Next, we use these reported rates to approximate $\rho(t)$ by assuming that the incidence rate is constant within each age-cohort $a_i$, i.e. $\rho(t)=\rho(a_i)=const$ for all $t \in a_i$.
Hence, we can approximate the time integral in Eq.~\eqref{eq:cum_risk_derivation} as:
\begin{equation}
    \int_0^t \rho (u) du\approx 5 \sum \rho(a_j), \quad \mbox{where} \quad j=0,5,\dots,i, \quad \mbox{and} \quad t\in a_i.
\end{equation}
to estimate $\mathcal{R}(t)$.

\subsubsection{Age-specific cumulative risk obtained from the model}
\label{sec:cum_risk_model}
Our complete cancer development model~\cite[Eq. (2.10)]{main_paper} does not have an explicit age structure, making a direct link to incidence data difficult.
However, it provides the  age-specific cumulative risk $\hat{\mathcal{R}}(t)$ by simulating a cohort of $N$ individuals up to life expectancy.
Note that for our complete cancer development model~\cite[Eq. (2.10)]{main_paper} the risk of cancer at age-band $a_i$ is given as:
\begin{equation*}
    \varrho(a_i) = \frac{c(a_i)}{N-\sum_{j<i} c(a_j)},\text{ and }j=0,5,\dots,i,
\end{equation*}
where $c(a_i)$ are realisations that develop cancer in the age-band $a_i$, and $N$ is the size of the simulated cohort.
Accordingly, age-specific cumulative risk
$\hat{\mathcal{R}}(t)$ with $t\in a_i$, is given as
\begin{equation*}
   \hat{\mathcal{R}}(t) = 1-\prod_{j\leq i} \left(1-\varrho\left(a_j\right)\right), ~~ t \in a_i.
\end{equation*}

\subsubsection{Technical details on simulations of the complete model}
\label{sec:tech_detail}
In this section, we provide further technical details about the underlying calculations performed in~\cite[Sec.3]{main_paper}.
We

\begin{itemize}
    \item Use a low-level implementation for the Monte-Carlo simulations of \cite[Eq. (2.10)]{main_paper}\footnote{To speed up the simulations, we approximate the binomial distribution with a truncated Poisson distribution with $\lambda(t)=n p(t)$.} either in Cuda 12.0 (wrapped by PyTorch's Cuda wrapper~\cite{PyTorch}) or a Numba~\cite{numba} compiled function.
    \item 
    Introduce a {\em detection threshold}
    $x_{d} = 3.05\times 10^9$  cells, corresponding to a tumour of $1.8$ cm diameter\footnote{This is based on the average detection size for biennial mammograms~\cite{average_detection_size}, since only a biennial mammography is provided free in the Irish health system~\cite{HSE_breast_screening}.
    We assume a sphere and convert the tumour volume to the number of cancer cells, assuming a cancer cell density of $10^9$ cells/cm$^3$~\cite{1cm3_cell_count}.
    }.
    We use $t_d$ to denote the time when a tumor reaches this size in a realization,  that is $x(t_d) = x_d$.  We then say the person is diagnosed with breast cancer at time $t=t_d$.
\end{itemize}

\subsubsection{Typical cumulative risk of most cancers: Fixed immune response}
\label{sec:fixed_immune}

\paragraph{Details on colorectal cancer}{
In the case of colorectal cancer, we choose the  detection threshold $x_d$ to be a  tumour of diameter of 2 cm~\cite{colonoscopy_guidelines}  and convert the tumour volume to the number of cancer cells, assuming a spherical tumour shape and cancer cell density of $10^9$ cells/cm$^3$~\cite{1cm3_cell_count}}, i.e. $x_d=4.2\times10^9$ cells.
The (black) cumulative risk $\mathcal{R}(t)$ of colorectal cancer (CRC) for females,  shown in Fig.~\ref{fig:crc}, is based on age-specific incidence rates averaged over the period from 1994 to 2021~\cite{NCRI}.
Specifically, the incidence shown for females within each age cohort $a_i = [i,  i+5) \text{ for }i=0, 5, \dots, 75,$  represents the average incidence for each age cohort $a_i$ over the period 1994 to 2021.
In particular, we selected in~\cite{NCRI}, ``By age" in the header, ``Colorectal|C18-C21" under Cancer, ``Ireland" under Region, ``Females" under Sex, ``1994" under Year from, and ``2021" under Year to.

\subsubsection{Modelling hormone replacement therapy}
\label{sec:HRT}
Based on observations by~\cite{hrt_Gamma_shape}, we model the duration of the hormone replacement therapy $H$ in years,  as a gamma distribution  with probability density function
\begin{equation}
    f(H) = \frac{1}{\Gamma(k) \theta^k}H^{k-1}e^{-H/\theta},
    \label{eq:gamma}
\end{equation}
with shape parameter $k$, scaling parameter $\theta$, and $\Gamma(k)$ denoting the standard gamma function.
The parameters are selected such that the mode and the standard deviation of \eqref{eq:gamma}.
The mode $M_\Gamma$ and the standard deviation $\sigma_\Gamma$ of the probability distribution in Eq.~\eqref{eq:gamma} are given as
\begin{equation}
    \begin{split}
        M_\Gamma &= (k -1)\theta,\\
        \sigma_\Gamma &= \sqrt{k \theta^2}.
    \end{split}
\end{equation}
We rearrange the above expressions to solve for the shape and scaling parameters:
\begin{equation}
    \begin{split}
        k &= 1+\frac{M_\Gamma \left(M_\Gamma +\sqrt{M_\Gamma^2 + 4 \sigma_\Gamma }\right)}{2 \sigma_\Gamma^2},\\
        \theta &= \frac{-M_\Gamma+\sqrt{M_\Gamma^2+4 \sigma_\Gamma^2}}{2},
    \end{split}
\end{equation}
and omit the solution that yields $\theta<0$.
Finally, we obtain $k\approx3.2553$ and $\theta\approx2.66$, by using $M_\Gamma = 6$ years with standard deviation $\sigma_\Gamma=4.8$ years~\cite{6_0_mean_HRT}.

\subsubsection{Effects of menstrual cycle length on breast cancer
risk}
\label{sec:cycle_risk_par_expl}
To investigate the relationship between risk and the length of the menstrual cycle, we performed Monte Carlo simulations for different cycle lengths.
In particular, for each follicular phase with length $t_f= 8,9,\dots,20$, we performed $N_{t_{f}}= 1\times 10^7$ realisations and run {\em Case 3} of \cite[Sec. 3 b) ii)]{main_paper} until the average menopause of 51 years.
First, we introduce the risk of developing cancer by the age of 51 years depending on the cycle length
\begin{equation}
    \hat{\mathcal{R}}(51,t_f) = \frac{c_{t_{f}}}{N_{t_{f}}} \text{ with }t_f = 8,9,\dots,20,
\end{equation}
where $c_{t_{f}}$ denotes the cases occurring for follicular phase length $t_f$.

Second, we introduce the average risk among a group of women whose follicular phase $t_f\in [a,b]$ 
\begin{align*}
    \bar{\mathcal{R}}_{a,b}=&\frac{1}{b-a}\int_a^b\hat{\mathcal{R}}(51,t_f)\;dt_f.
\end{align*}
In~\cite{nearly_double_risk_short_cycle}, various risk factors were compared, including the length of the menstrual cycle.
They compared the average risk of women with a cycle length between 25 and 31 days to women with a cycle shorter than 25 days.

Our model estimates the risk for women with a cycle between 25 and 31 days, i.e. $t_f\in[11,17]$ to be
\begin{equation}
    \bar{\mathcal{R}}_{11,17} \approx 0.0295 = 2.95\%,
\end{equation}
using the trapezoidal rule for integration. 
To estimate the risk for women with a cycle shorter than 25 days, we need to choose a sensible lower bound.
Recall that we assumed the cycle length is taken from a normal distribution with mean 28 days and standard deviation of 2.4 days.
Thus, a lower bound of 22 days, i.e. $t_f=8$, lies at the 98\% confidence interval.
In other words, the lower bound is chosen to be approximately 2 standard deviations below the mean.
Therefore, we estimate the risk for women with a cycle length between 22 and 25 days, i.e. $t_f\in[8,11]$, to be
\begin{equation}
    \bar{\mathcal{R}}_{8,11}\approx 0.0517 = 5.17\%.
\end{equation}
As relative risk between both groups we obtain
\begin{equation}
    rr = \frac{\bar{\mathcal{R}}_{11,17}}{\bar{\mathcal{R}}_{8,11} } \approx 1.7525.
\end{equation}
which is in a very good agreement to the value 1.86 reported in~\cite{nearly_double_risk_short_cycle}.
Note that the relative risk increases further if even shorter cycles are considered.

\section{Breast cancer progression in  women}

Data set 3 reported by Förnvik {\em et al.}~\cite{ct_scans_breast_bvp} contains two measurements of tumour diameter taken half a year apart in one human female.
The initial diameter is 10mm and after six months it is 24mm.
\footnote{We could not find more detailed data of untreated human breast cancer progression in the literature.}
Tumour diameters are converted into number of cancer cells assuming a spherical shape and a cell density of $10^9$ cell/cm$^3$~\cite{1cm3_cell_count}, to facilitate model fitting.

\begin{figure}[th]
    \centering
         \includegraphics[width=0.7\textwidth]{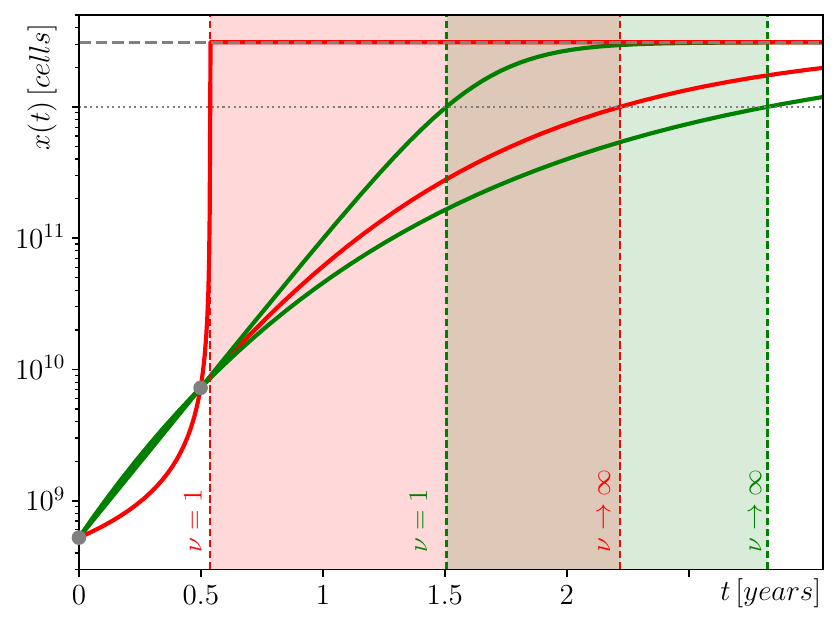}
    \caption{
    (Grey dots) The two measurements reported in~\cite{ct_scans_breast_bvp}, together with the best fits for (red curves) Model 1 and (green curves) Model 2 for the two values of the growth parameter $\nu=1$ and $\nu\to\infty$. The survival time for (red band) Model 1 ranges from 0.5 year to 2.2 years, and (green band) for Model 2 from 1.51 to 2.82 years.
    The estimated values for all parameters are listed in Tab.~\ref{tab:parameter_table_human_estimates}.
    }
    \label{fig:BVP_human_rh_volt}
\end{figure}

Given the paucity of data, certain parameter values are chosen guided by~\cite{Norton_Breast_K}.
We fix the carrying capacity at $K=3.1\times 10^{12}$ cancer cells, which corresponds to a tumour diameter of 181 mm or a tumour volume of 3.1 litres, indicated by the dashed horizontal line in Fig.~\ref{fig:BVP_human_rh_volt}.
We also fix  the lethal tumour size at $x_l =10^{12}$ cancer cells, which corresponds to a tumour diameter of 124 mm or a tumour volume of 1 litre, indicated by the dotted horizontal line in Fig.~\ref{fig:BVP_human_rh_volt}.
Additionally, we assume that the extinction threshold is one order of magnitude smaller than the current tumour detection threshold.
The detection threshold is 1mm diameter~\cite{1mm_mammo}, or $5.24\times10^5$ cancer cells giving an extinction threshold of $A= 10^4$ cancer cells, corresponding to a tumour diameter of $\approx 0.3 mm$.
The remaining model parameters are estimated by requiring that the solution passes through both measured data points.
Our results are given in terms of patient survival time, the time it takes the tumour to reach the lethal threshold.

We begin with Model 1 in its factorised form given in Eq.~\eqref{eq:gen_Allee_app}.
We consider two special cases: Volterra's model with $\nu=1$, and Gompertz's model with an extinction threshold for $\nu\to\infty$ given by Eq.~\eqref{eq:limit_volterra}.
This leaves only one parameter, $r$, to be estimated.
We determine $r$ by solving the following boundary value problem
\begin{equation*}
    \dot{x} = f(x; r),
\end{equation*}
with boundary conditions: 
\begin{equation}
    \begin{split}
        0&= x(t_1)- x_1,\\
        0&= x(t_2)- x_2,\\
    \end{split}
\end{equation}
using a shooting method.
We use the previously introduced measurements by Förnvik {\em et al.}~\cite{ct_scans_breast_bvp}, i.e. $x_1\approx 5.245\times10^8$ cells, $x_2\approx7.24\times 10^9$ cells, $t_1=0$ days, and $t_2=182$ days.
For $\nu=1$, the best fit is a step-like (red) growth curve in Fig.~\ref{fig:BVP_human_rh_volt}, and a survival time of 0.5 years (196 days).
Taking $\nu\to\infty$ the best fit is a smoother (red) growth curve, and a survival time of 2.2 years.

Next, we consider Model 2 in~\eqref{eq:richards_harvest_appendix} with four parameters $\nu, r, \mu$ and $s$  to be determined.
We constrain the values of $A$ and $K$ to be the same as those used in Model 1 so that the models can be compared.
We set up the following boundary value problem
\begin{equation*}
    \dot{x} = f(x; r,\nu, \mu, s),
\end{equation*}
with boundary conditions: 
\begin{equation}
    \begin{split}
        0 &= x(t_1)- x_1,\\
        0 &= x(t_2)-x_2,\\
        0 &= f(x; r,\nu, \mu, s)\vert_{x=A},\\
        0 &= f(x; r,\nu, \mu, s)\vert_{x=K},
    \end{split}
    \label{eq:boundary_conditions}
\end{equation}
Two special cases, $\nu=1$ and $\nu\to\infty$, are again considered.
For $\nu=1$, the best fit is a smooth (green) growth curve in Fig.~\ref{fig:BVP_human_rh_volt}, with a survival time of 1.51 years.
On the other hand, for $\nu\to\infty$, the best fit is a (green) growth curve with survival time of 2.82 years. 

Model 1 encapsulates a wider range of survival times than Model 2. 
On the other hand, Model 2 overlaps with reported historical untreated survival times for breast cancer, i.e. 2.3 to 2.7 years~\cite{untreated_breast_cancer}. 
We conclude that Model 2 is most appropriate for modelling cancer progression.
All model parameters that give the best fit are listed in Table~\ref{tab:parameter_table_human_estimates}.

The parameter values used in the complete cancer development model~\cite[Eq. (2.10)]{main_paper} and reported in Table \ref{tab:parameter_table_human}, were obtained using $\nu=10$ and solving the boundary value problem given above in Eq.~\ref{eq:boundary_conditions}.

\section{Functional form of $\mathcal{R}(t)$ for various cancer types}
\begin{figure*}
    \centering
    \includegraphics[width=\textwidth]{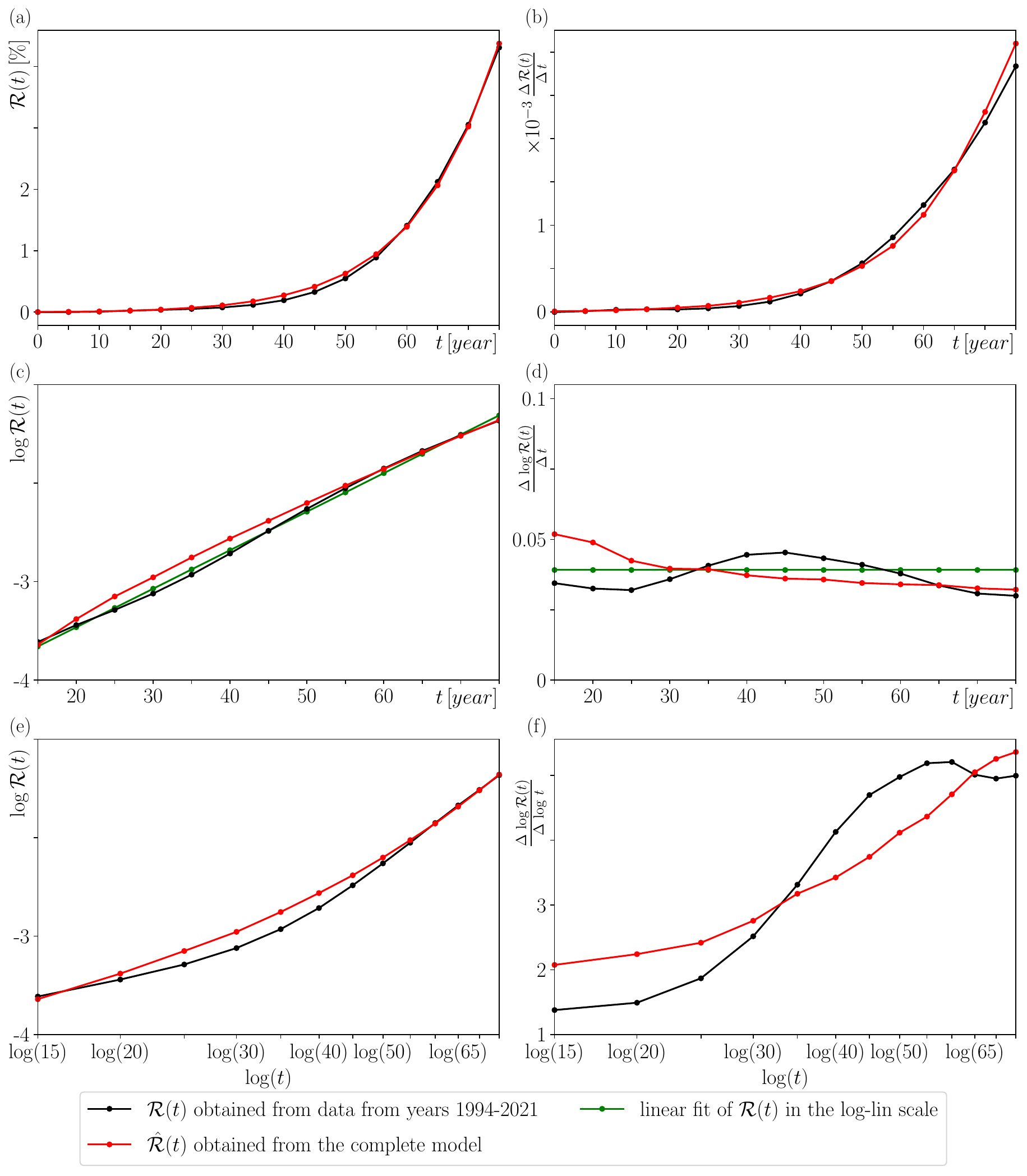}
    \caption{
    The age-specific cumulative risk of colorectal cancer in women $\mathcal{R}(t)$ represented on a lin-lin (a), log-lin (c), and log-log (e) scale.
    Approximate values of derivatives, using central differences, are shown in (b), (d), and (f) respectively.
    The (black) risk $\mathcal{R}(t)$ is based on averaged data from the years 1994 to 2021 obtained by the NCRI~\cite{NCRI}.
    The (red) risk $\mathcal{R}(t)$, is the output from the complete cancer development model \cite[Eq. (2.10)]{main_paper} with a constant immune response $s_m(t) = s_{max} = const$ as depicted in \cite[Fig.~6]{main_paper} and parameter values given in the caption.
    The age-specific risk increases exponentially with age at a (green) rate of 0.039 [$\text{year}^{-1}$] because there is linear relationship  evident in (c) and almost constant derivative in (d).
    Furthermore, the relationship in (e) is non-linear on a log-log plot indicating that a power law form does not apply.
    }
    \label{fig:crc}
\end{figure*}

\begin{figure*}
    \centering
    \includegraphics[width=\textwidth]{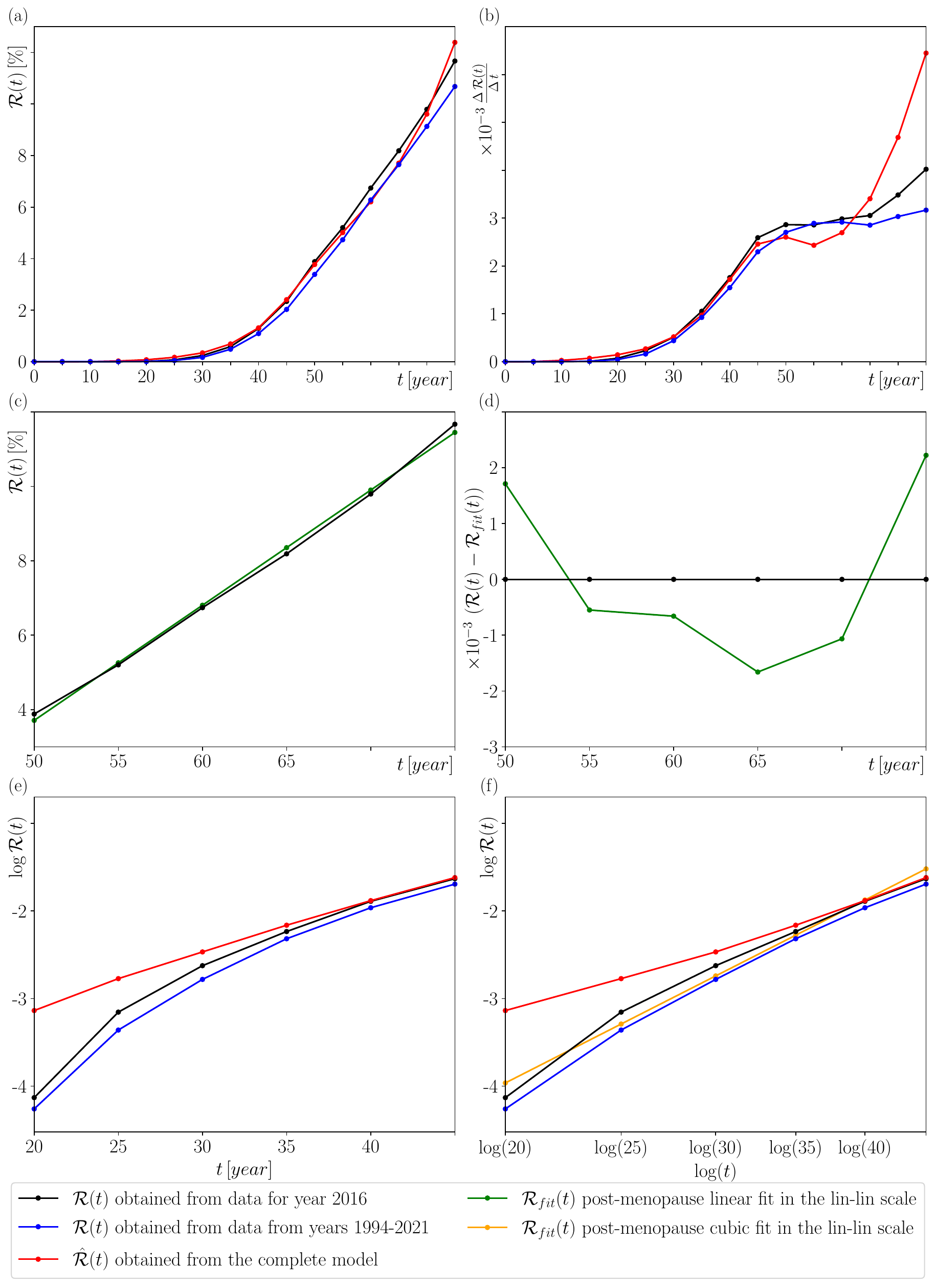}
    \caption{
    The shape of the age-specific risk of breast cancer $\mathcal{R}(t)$ represented on a lin-lin (a), (c), (d), a log-lin (e), and log-log scale (f).
    Approximate values of derivatives $\mathcal{R}(t)$ represented on a lin-lin scale, using central differences, are shown in (b).
    The (blue) risk $\mathcal{R}(t)$ is based on averaged data from the years 1994 to 2021 appears 
    The (black) risk $\mathcal{R}(t)$ is based on data of the census year 2016 which the (red) risk $\mathcal{R}(t)$, based on the complete cancer development model \cite[Eq. (2.10)]{main_paper} and case 3 of \cite[Sec. 3 b) ii)]{main_paper}, is fitted to.
    Pre menopause (up to the age of 45-50), the age-specific cumulative risk of breast cancer in women is a 7th order polynomial as evidenced by the (orange) linear fit $\mathcal{R}(t)$ in (f) with slope $a\approx6.94$ and the nonlinear scale in the log-lin plot (e).
    Post menopause (after the age 50), becomes linear as evidences by a (purple) cubic fit with vanishing parameters of the cubic and quadratic term in (c) and residuals in (d) (see Tab.~\ref{tab:poly_parameter} for parameters).
    The slope is obtained by a (green) linear fit in (c) and (c) (see Tab.~\ref{tab:poly_parameter} for parameters).
    }
    \label{fig:breast}
\end{figure*}

\begin{figure*}
    \centering
    \includegraphics[width=\textwidth]{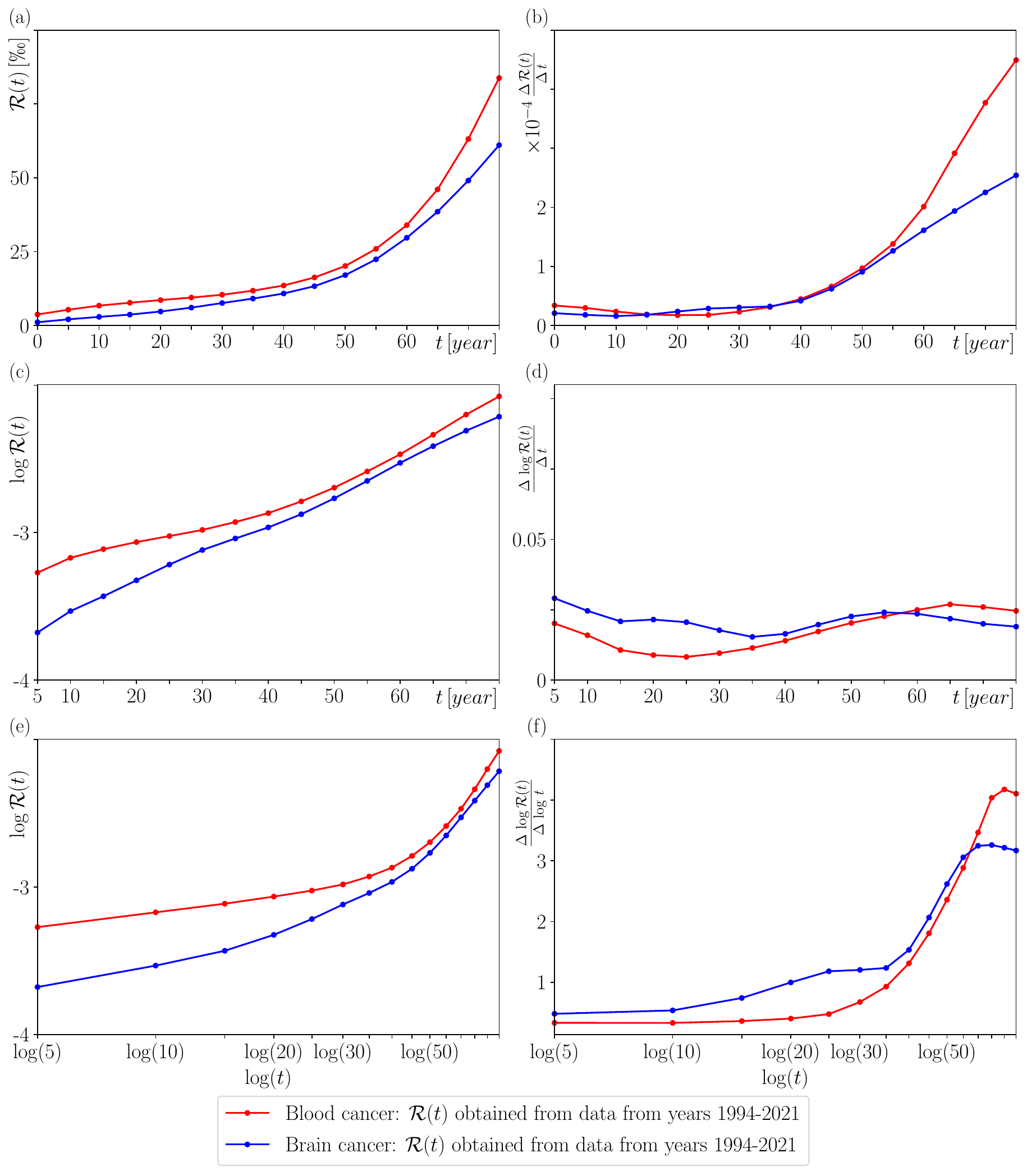}
    \caption{
    The age-specific cumulative risk of (red) blood and (blue) brain cancer in women $\mathcal{R}(t)$ represented on a lin-lin (a), log-lin (c), and log-log (e) scale.
    The approximated derivatives are given in (b), (d), and (f) respectively.
    Both (blue and red) risks $\mathcal{R}(t)$ are based on averaged data from the years 1994 to 2021 obtained by the NCRI~\cite{NCRI}.
    Both cancer types appear to increase exponentially with age, according to the linear representation in (c) confirmed by a nearly constant derivative (d) and a nonlinear representation in (e).
    }
    \label{fig:otherTypes}
\end{figure*}

As highlighted by~\cite{aging_stem_cell_driver} age is one of the strongest predictors of cancer incidence~\cite{age_cancer_predictor}.
It is commonly observed that incidence (and therefore the cumulative risk) of many cancer types, such as colorectal cancer (CRC), brain, blood and even breast cancer increase with age as shown in Fig.~\ref{fig:crc} (a), Fig.~\ref{fig:otherTypes} (a) and Fig.~\ref{fig:breast} (a).\footnote{An exception would be testicular cancer, which is most common for young men (15-34 years)~\cite{testis_cancer}.}

In this section we aim to determine the functional form of $\mathcal{R}(t)$, the cumulative risk at age $t$, for brain, blood and breast cancer  in women.
First, we consider an exponential form:
\begin{equation}
    \mathcal{R}(t) = a b ^t,
    \label{eq:exp_func_form}
\end{equation}
and take the logarithms to obtain
\begin{equation*}
    \log \mathcal{R}(t) =  \log (a) + \log(b) t.
\end{equation*}
If we plot $\log \mathcal{R}(t)$ versus $t$ (semi-logarithmic or lin-log plot), we expect a straight line with slope $\log(b)$ if $\mathcal{R}(t)$ follows an exponential form in Eq.~\eqref{eq:exp_func_form}.
We also consider a power law form:
\begin{equation}
    \mathcal{R}(t) = a t^ b
    \label{eq:power_law}
\end{equation}
and take the logarithms to obtain
\begin{equation*}
    \log \mathcal{R}(t) =  \log (a) + b\log(t).
\end{equation*}
If we plot $\log \mathcal{R}(t)$ versus $\log(t)$ (logarithmic or log-log plot),  we expect a straight line with slope $b$ if  $\mathcal{R}(t)$ follows a power law form in Eq.~\eqref{eq:power_law}.

In Fig.~\ref{fig:crc} (c), we see that the  (black)  actual age-specific cumulative risk of colorectal cancer in women is linear on a lin-log plot.
This strongly indicates that the relationship between cumulative risk of colorectal cancer in women and age is exponential with rate $3.911447\times10^{-2}$ $\text{year}^{-1}$.

Furthermore, the age-specific cumulative risk of  (red) blood and (blue) brain cancer in women also demonstrates an exponential relationship because there is a linear relationship on a lin-log plot (see Fig.~\ref{fig:otherTypes} (c)).

In contrast to the results for colorectal, blood and brain cancer, the age-specific cumulative risk of breast cancer is significantly different (see Fig.~\ref{fig:breast} (a) and (b)).
In particular, we note that after age 50 the  (black) actual cumulative risk increases approximately linearly (Fig.~\ref{fig:breast} (a)) and the derivative (Fig.~\ref{fig:breast} (b)) flattens. In Fig.~\ref{fig:breast} (c) a straight line is fit to the 50 and over portion of the cumulative risk curve
\begin{equation*}
   \mathcal{R}(t) = ct+d,
\end{equation*}
with obtained parameter values given in Tab.~\ref{tab:poly_parameter}.
We also fit a cubic polynomial to the curve and it has equation
\begin{equation*}
   \mathcal{R}(t) = a t^3+b t^2+ct+d.
\end{equation*}
The coefficients of the cubic and quadratic are 4 and 2 orders of magnitude smaller than the linear coefficient (see Tab.~\ref{tab:poly_parameter}).
This adds to our confidence that a linear trend is present.

On the other hand, analysing the  (black)  actual data for the age cohort 20 to 45 (see Fig.~\ref{fig:breast} (e) and (f)), we conclude that a power law relationship is most appropriate because the relationship tends to be linear on a log-log plot.
The best fit line gives power law parameters of $b = 7$  and $a = 6.93$.

\clearpage
\section{Tables}

\begin{table}[th]
    \caption[]
    {
    These following parameter values for Model 1 and 2 are obtained through fitting both to the data sets of Vaghi {\em et al.}~\cite{mice_data_cloud} and Cabeza {\em et al.}~\cite{treatment_mice} and depicted in \cite[Fig. 5]{main_paper}.
    }
    \centering
    \begin{tabular}{cccc}
    \toprule
    parameter & unit & Vaghi {\em et al.}~\cite{mice_data_cloud} & Cabeza {\em et al.}~~\cite{treatment_mice}  \\
    \midrule
    \multicolumn{4}{c}{Model 1 $\nu=1$} \\
    \midrule
    $\nu$   &  & 1                         & 1                 \\
    $r$     & 1/days  & -0.04269035          & $-0.00271811\times10^{-4}$      \\
    $A$     & cells   & 402434.641            & 31712.9110        \\
    $K$     & cells   & $7.2049\times 10^{8}$ & $12.2290\times 10^{9}$        \\
    \midrule
    \multicolumn{4}{c}{Model 1 $\nu$ free} \\
    \midrule
    $\nu$   &  & 77799.4520                 & 6.05194241        \\
    $r$     & 1/days & -0.00543638    & $-0.00297248\times10^{-4}$\\
    $A$     & cells & 10.0000576      & $199.822097$ \\
    $K$     & cells & $1.9696\times 10^{9}$           & $3.2734\times 10^{9}$       \\
    \midrule
    \multicolumn{4}{c}{Model 2}\\
    \midrule
    $\nu$   &       & 97196        & 1                  \\
    $r$     & 1/days & 0.07496541         & 0.36148949          \\
    $\mu$   & $\text{cell}^{-1/\nu}/\text{day}$ & 0.07494869  & $1.2084\times 10^{-10}$ \\
    $s$     & cells/days & 275.094444         & 108528.065        \\
    $A$     & cells & $225.66$        & $3.00254\times 10^5$         \\
    $K$     & cells & $2.60785342\times 10^9$         & $2.9911\times 10^9$     \\
    \bottomrule
    \end{tabular}
    \label{tab:parameter_table_mice_estimates}
\end{table}

\begin{table}[th]
    \centering
    \caption[]{The parameter values obtained through fitting (red) $\hat{\mathcal{R}}(t)$ to (black) $\mathcal{R}(t)$ and correspond to colorectal cancer (see \cite[Fig. 6]{main_paper}) and breast cancer (see \cite[Fig. 7]{main_paper}).}
    \begin{tabular}{ccccc}
        \toprule
         & $p_0$ & $p_T$ & $s_{min}$ [cells/day]& $s_{max}$ [cells/day] \\
         \midrule
         \cite[Fig. 6]{main_paper}   & $2.73943844 \times 10^{-5}$ & $2.74027764 \times 10^{-5}$ &  & 274250.0  \\
         \midrule
         \midrule
         \cite[Fig. 7 (a)]{main_paper}  & $2.73990516 \times 10^{-5}$ & $2.74106446 \times 10^{-5}$ & 274300.912 & 274300.912  \\
         \cite[Fig. 7 (b)]{main_paper}     & $2.73940502 \times 10^{-5}$ & $2.74080039 \times 10^{-5}$ & 274203 & 274295 \\ 
         \cite[Fig. 7 (c)]{main_paper}   & $2.73940501 \times 10^{-5}$ & $2.74080040 \times 10^{-5}$ & 274203 & 274295 \\ 
         \cite[Fig. 7 (d)]{main_paper}    & $2.73940451 \times 10^{-5}$ & $2.74079969 \times 10^{-5}$ & 274203.0 & 274295.0 \\
         \bottomrule
    \end{tabular}
    \label{tab:s_p_fit}
\end{table}

\begin{table}[th]
\caption{The parameter values used by the complete cancer development model~\cite[Eq. (2.10)]{main_paper} in \cite[Sec. 3]{main_paper}.}
\centering
\begin{tabularx}{0.95\textwidth}{cccX}
    \toprule
    parameter   & value                     & unit             & info \\
    \midrule
    $x(t)$      & $[0,\infty)$              & cells    & cell count at time $t$\\
    $A$         & $10^{4}$              & cells    & 
    Assumed to be smaller than $1$ mm diameter~\cite{1mm_mammo}\\
    $K$         & $3.1\times 10^{12}$           & cells    & Obtained from~\cite{Norton_Breast_K}.\\
    $t$         & $[0,T]$                   & days     & time\\
    $T$         & $29200$                   & days       & Corresponding to 80 years and neglecting leap years \\
    \midrule
    \multicolumn{4}{c}{Model 2}\\
    \midrule
    $r$         & $2.78777026\times10^{-3}$   & 1/days   & Obtained through fit to~\cite{ct_scans_breast_bvp}\\
    $\mu$       & $1.57079970\times10^{-4}$   & $\text{cells}^{-1/\nu}/\text{day}$    & Obtained through fit to~\cite{ct_scans_breast_bvp}\\
    $\nu$       & $10$                      &                   & chosen\\
    $s$      & 239.32032                          & cells/days & Obtained through fit to~\cite{ct_scans_breast_bvp}\\
    \midrule
    \multicolumn{4}{c}{Complete cancer developement model}\\
    \midrule
    $s_{max}$   & 274203.0                & cells/day & Immune system strength {\bf without} progesterone\\
    $s_{min}$   & 274295.0                & cells/day & Immune system strength {\bf with} progesterone\\
    $n$     & $10^{10}$                   & cells      & Estimated number of stem cells inside the breast\\
    $p(t)$      & $[0,1]$                   &           & Probability of a single cell to mutate\\
    $p_0$      & $2.73940451 \times 10^{-5}$                   &           & Probability of a single cell to at birth\\
    $p_T$      & $2.74079969 \times 10^{-5}$                  &           & Probability of a single cell to at life expectancy\\
    $m(t)$      & $[0,\infty)$              & cells      & Number of mutations\\
    $l$         & $[8,20]$                  & days       & Length of the follicular phase\\
    $\sigma_l$  & $2.4$                    & years      & Variance of the follicular phase~\cite{var_cycle}.\\
    $\kappa$    & $12$                      & years      & Menarche; Obtained from~\cite{12_menarche}.\\
    $\mu_m$     & $51$                      & years      & Menopause mean; Obtained from~\cite{51_meno}.\\
    $\sigma_m$  & $4.86$                    & years      & Menopause SD; Obtained from~\cite{4_86_meno_sd}.\\
    $p_{hrt}$   & 0.26                          &           & Proportion of women that use HRT  \\
    $k_{hrt}$ & 3.2553                            & years      & Shape parameter of HRT-Gamma distribution\\
    $\theta_{hrt}$& 2.66                        & years      & Scale parameter of HRT-Gamma distribution\\
    \bottomrule
\end{tabularx}
\label{tab:parameter_table_human}
\end{table}

\begin{table}[]
    \caption[]
    {
    Below the common used values used for the human bvp and the remaining estimated values are listed.
    }
    \centering
    \begin{tabular}{cccc}
    \toprule
    parameter & units&\multicolumn{2}{c}{BVP Common values}\\
    \midrule
    $t_1$    & days  &\multicolumn{2}{c}{0}  \\
    $t_2$    & days  &\multicolumn{2}{c}{182}  \\
    $x_1$    & cells &\multicolumn{2}{c}{$5.235987755\times10^8$}\\
    $x_2$    & cells &\multicolumn{2}{c}{ $7.238229473\times 10^9$}\\
    $x_{l}$  & cells &\multicolumn{2}{c}{$10^{12}$}\\
    $A$      & cells & \multicolumn{2}{c}{$10^4$}         \\
    $K$      & cells & \multicolumn{2}{c}{$3.1\times10^{12}$}\\
    \midrule
    \midrule
     &       & Lower bound & Upper bound \\
    \midrule
    \multicolumn{4}{c}{Model 1} \\
    \midrule
    $\nu$   & 1/days & 1                   & $\infty$                 \\
    $r$     & 1/days & $-9.73939594\times10^{8}$    & -0.00016202       \\
    $t_{l}$ & days & 196.26   & 810.05        \\
    \midrule
    \multicolumn{4}{c}{Model 2}\\
    \midrule
    $\nu$   &                                   & 1                & $\infty$                  \\
    $r$     & 1/days                             & $1.44428169\times10^{-2}$   & $1.97826781\times10^{-3}$        \\
    $\mu$   & $\text{cells}^{-1/\nu}/\text{days}$ & $4.65897317\times10^{-15}$   & $1.97826212\times10^{-3}$  \\
    $s$     & cells/days                          & 1.44428168       & $3.86792184\times10^{2}$        \\
    $t_{l}$ & days & 550         & 1030         \\
    \bottomrule
    \end{tabular}
    \label{tab:parameter_table_human_estimates}
\end{table}

\begin{table}
    \caption{The parameter values to determine the functional form of the cumulative risk of breast cancer $\mathcal{R}(t)$ are listed below.}
    \centering
    \begin{tabular}{ccccc}
        \toprule
         & a & b & c & d\\\midrule
        \midrule
        \multicolumn{5}{c}{Colorectal cancer (see Fig.~\ref{fig:crc})}\\
        \midrule
        Exponential & $3.911447\times10^{-2}$  & -4.24587996 & &\\
        \midrule
        \multicolumn{5}{c}{Breast cancer (see Fig.~\ref{fig:breast})}\\
        \midrule
        \midrule
         Linear   & 0 & 0 & $3.09493\times10^{-3}$  & $-1.1764878\times10^{-1}$ \\
         Cubic    & $7.5313\times10^{-7}$ & $-1.1938\times10^{-4}$ & $9.096\times10^{-3}$ & $-2.118687\times10^{-1}$\\
        \midrule
        Power law& 6.93898337 & -12.99075653 & &\\
        \bottomrule
    \end{tabular}
    \label{tab:poly_parameter}
\end{table}

\clearpage
\bibliographystyle{unsrt}
\bibliography{references.bib}